\documentclass[12pt, leqno]{article}
\usepackage{amsmath, setspace, multirow, lineno, ulem}
\usepackage[font=small, labelfont=bf, singlelinecheck=off]{caption}
\usepackage[top=1.5in, bottom=1.5in, left=1.1in, right=1.1in]{geometry}
\usepackage[authoryear,round]{natbib}
\usepackage{color,soul}

\DeclareCaptionStyle{italic}[justification=centering]{labelfont={bf}, textfont={it}, labelsep=colon}

\usepackage{graphicx, psfrag, epsf, textcomp, epstopdf, amsthm, paralist}
\usepackage{enumerate, dsfont, alltt, verbatim, stackengine, scalerel}
\usepackage{url,xcolor}
\usepackage{booktabs, subfig, bm, paralist, mathpazo, tikz, longtable, microtype, authblk}
\usepackage[linesnumbered, ruled, vlined]{algorithm2e}
\stackMath            
            
\definecolor{darkblue}{rgb}{0,0,.6}
\definecolor{DarkRed}{rgb}{.7,0,.4}

\usepackage[pdftex,colorlinks=true,citecolor=darkblue,linkcolor=darkblue,urlcolor=darkblue,pdfborder={0 0 0}]{hyperref}

\usepackage{comment,orcidlink,doi}

\newcommand{\blind}{0}

\newcommand{\X}{\mathcal{X}}
\newcommand{\Y}{\mathcal{Y}}

\newcommand{\Rlogo}{\protect\includegraphics[height=1.8ex,keepaspectratio]{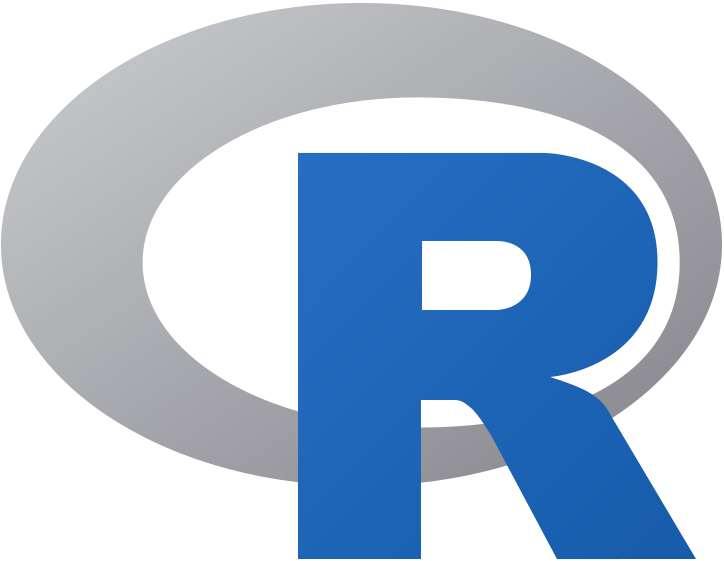}}

\graphicspath{{plots/}}

\newsavebox\CBox

\newtheorem{@definition}{\sc Definition}[section]

\newtheorem{theorem}{\sc Theorem}[section]
\newtheorem{remark}{\sc Remark}[section]

\renewcommand\X{\mathcal{X}}

\date{}

\begin{document}

\def\spacingset#1{\renewcommand{\baselinestretch}{#1}\small\normalsize} \spacingset{1}

\if0\blind
{
\title{\bf Penalized spatial function-on-function regression}

\author[1]{\normalsize Ufuk Beyaztas\thanks{Corresponding address: Department of Statistics, Marmara University, 34722, Kadikoy-Istanbul, Turkey; Email: ufuk.beyaztas@marmara.edu.tr} \orcidlink{0000-0002-5208-4950}
}
\author[2]{\normalsize \quad Han Lin Shang \orcidlink{0000-0003-1769-6430}}
\author[1]{,\normalsize \quad Gizel Bakicierler Sezer \orcidlink{0000-0002-1789-0842}}

\affil[1]{\normalsize Department of Statistics, Marmara University, Turkey}
\affil[2]{\normalsize Department of Actuarial Studies and Business Analytics, Macquarie University, Australia}
}

\maketitle
\fi

\if1\blind
{
\title{\bf Penalized spatial function-on-function regression}
}
\fi

\maketitle

\begin{abstract}
The function-on-function regression model is fundamental for analyzing relationships between functional covariates and responses. However, most existing function-on-function regression methodologies assume independence between observations, which is often unrealistic for spatially structured functional data. We propose a novel penalized spatial function-on-function regression model to address this limitation. Our approach extends the generalized spatial two-stage least-squares estimator to functional data, while incorporating a roughness penalty on the regression coefficient function using a tensor product of B-splines. This penalization ensures optimal smoothness, mitigating overfitting, and improving interpretability. The proposed penalized spatial two-stage least-squares estimator effectively accounts for spatial dependencies, significantly improving estimation accuracy and predictive performance. We establish the asymptotic properties of our estimator, proving its $\sqrt{n}$-consistency and asymptotic normality under mild regularity conditions. Extensive Monte Carlo simulations demonstrate the superiority of our method over existing non-penalized estimators, particularly under moderate to strong spatial dependence. In addition, an application to North Dakota weather data illustrates the practical utility of our approach in modeling spatially correlated meteorological variables. Our findings highlight the critical role of penalization in enhancing robustness and efficiency in spatial function-on-function regression models. To implement our method we used the \texttt{robflreg} \Rlogo \ package on CRAN.
\end{abstract}


\newpage
\spacingset{1.65} 

\section{Introduction} \label{sec:1}

The function-on-function regression (FoFR) model proposed by \cite{ramsay1991} has emerged as a paradigm of choice to capture the intricate relationships between functional covariates and responses. The FoFR model is particularly valuable for modeling continuous phenomena that evolve over time or space, addressing challenges inherent in high-dimensional, dependent data with remarkable precision and insight. The model has undergone significant methodological advances, cementing its role as a versatile and indispensable tool in modern data analysis \citep[see, e.g.,][]{yao2005, MullerYao2008, wang2014, ivanescu2015, BS20, BS20C, Cai2022, wang2022}.

To formulate the FoFR model, let $(\Y_i,\X_i)$, $i \in \{1, \ldots, n \}$, denote an independent and identically distributed (i.i.d.) sample from the joint distribution of $(\Y,\X)$, where $\Y = \Y(t)$ and $\X = \X(s)$ represent the functional response and functional covariate, respectively. The response $\Y$ is defined in the compact domain $t \in \mathcal{I}_y \subset \mathbb{R}$, while the covariate $\X$ is defined in $s \in \mathcal{I}_x \subset \mathbb{R}$. Accordingly, $\Y_i(t)$ and $\X_i(s)$ are realizations of stochastic processes, mapping $\mathcal{I}_y \rightarrow \mathbb{R}$ and $\mathcal{I}_x \rightarrow \mathbb{R}$, respectively. The FoFR model is given by:
\begin{equation*}
\Y_i(t) = \beta_0(t) + \int_{\mathcal{I}_x} \X_i(s) \beta(t,s) \, ds + \epsilon_i(t),
\end{equation*}
where $\beta_0(t)$ represents the intercept function, $\beta(t,s)$ is the bivariate regression coefficient function that captures the dynamic relationship between $\Y_i(t)$ and $\X_i(s)$, and $\epsilon_i(t)$ is the stochastic error process. The error term satisfies $\mathbb{E}\{\epsilon(t)\} = 0$ and $\mathrm{Var} \{\epsilon(t)\} < \infty$ for all $t \in \mathcal{I}_y$ that ensure the well-posedness of the model.

Although the FoFR model has been extended to encompass nonlinear relationships \citep{Lian2007, Schepl2015, Kim2018, BSQuad, rao2023, BSM2024} and nonparametric formulations \citep{wang2019, wang2022JCG, Boumahdi2023}, the literature predominantly assumes independence among observations. This simplifying assumption, though mathematically convenient, is often incongruent with the realities of data exhibiting inherent spatial dependence, a pervasile characteristic in disciplines such as public health, environmental monitoring, and geostatistics. In such contexts, overlooking spatial correlation compromises the validity of parameter estimates and undermines the robustness of inferential procedures, potentially leading to misleading scientific conclusions and flawed policy recommendations.

In the realm of spatially-correlated functional data, significant advancements have been made through kriging-based methodologies tailored for point-referenced data \citep{Nerini2010, Giraldo2011, Zhang2011, Caballero2013, Zhang2016, Bohorquez2016, Menafoglio2017, Aguilera2017, Bohorquez2017, Giraldo2018}. Despite these developments, exploring spatial dependence in functional responses for areal data has been relatively limited. In particular, the FoFR framework remains largely underdeveloped in addressing spatially structured dependencies within areal functional data, presenting a critical gap in the methodological literature and an opportunity for further advancements in the field.

In spatial regression modeling, three canonical formulations are commonly acknowledged: the spatial autoregressive model, the spatial error model, and the spatial Durbin model \citep{Lesage2009}. Of these, the spatial autoregressive model stands out for its ability to capture spatial dependence by including a spatially lagged response variable. This parsimonious representation succinctly encapsulates spatial interactions and offers significant advantages in terms of theoretical generalizability and computational efficiency \citep{Huang2021}. These attributes make the spatial autoregressive model particularly well suited for integration into the FoFR framework, providing a robust foundation for extending FoFR methodologies to accommodate spatially dependent functional data.

Despite these advancements, the development of methodologies for spatially dependent functional responses with functional predictors remains in its infancy. Existing studies have largely addressed specific and restricted scenarios, such as time series-based social network models that disregard temporal interactions between time points \citep[see, e.g.,][]{Zhu2022}, or autoregressive models for functional responses that incorporate only scalar covariates \citep[see, e.g.,][]{Hoshino2024}. However, the absence of a unified and comprehensive approach capable of simultaneously estimating spatial lag parameters and bivariate regression coefficient functions within the FoFR framework represents a significant gap in the literature.

Recently, \cite{BSGARC2024} introduced a spatial FoFR (SFoFR) model, providing a foundational framework for incorporating spatial dependencies in functional regression. Their approach employs a spatial functional principal component decomposition for the response and predictor, which transforms the infinite-dimensional SFoFR model into a finite-dimensional multivariate spatial autoregressive framework based on the principal component scores. The parameters of this finite-dimensional model are then estimated using a least-squares method. While this is an important step, the spatial functional principal component-based methodology has notable limitations that may hinder its practical application. 

First, estimating the number of principal components to retain is a challenging task, as the addition or omission of components can substantially alter the shape and interpretation of the estimated functional parameters \citep[see, e.g.,][]{Crainiceanu2009, ivanescu2015}. Second, the smoothness of the estimated coefficient functions is inherently tied to the smoothness of the functional principal component eigenfunctions and the number of functional principal components selected. This indirect control can lead to suboptimal smoothing, especially when the true functional parameters are considerably smoother than the higher-order principal components \citep[see, e.g.,][]{ivanescu2015}.

To address these challenges, we propose a penalized SFoFR (``PSFoFR'') that avoids the pitfalls of principal component truncation. We incorporate a penalized spatial two-stage least-squares estimator (PenS2SLS) that treats the spatial-lag term as endogenous while controling smoothness through data-driven penalties. In effect, rather than transforming the problem, we represent both the spatial kernel $\rho(t,u)$ (see~\ref{eq:sfofrm}) and the regression surface $\beta(t,s)$ with tensor-product B-splines and estimate them within a penalized FoFR framework, yielding stable fits with principled roughness regularization. The endogeneity of the spatially lagged functional response is handled by casting estimation as a generalized spatial two-stage least-squares problem and using spatial-lag instruments, extending the \cite{Kelejian1998} instrumental-variables (IV) lineage to functional parameters. This directly targets identification in spatial autoregressive-type models while retaining explicit smoothness control.

Nonparametric penalization for FoFR is well established in the mixed-model/penalized-spline literature. In particular, the penalized FoFR (pffr) of \cite{ivanescu2015} treats the coefficient surfaces as smooth random effects and selects smoothing parameters within a generalized additive model framework; this affords flexible surface estimation and approximate interval estimation, but is developed under independent errors and without spatial-lag endogeneity in the response. \cite{Aguilera2017} address spatially correlated functional data through a three-dimensional P-spline penalty (space–space–time) primarily for prediction, again without an endogenous spatial lag of the functional response. These two strands—pffr via mixed models and penalized spatial smoothing for functional prediction—thus target complementary problems: smooth inference for FoFR under independence \citep{ivanescu2015} versus spatial interpolation/smoothing of functional fields \citep{Aguilera2017}. 

The proposed PSFoFR estimator is built around a PenS2SLS procedure: we represent both the spatial kernel $\rho(t,u)$ and the regression surface $\beta(t,s)$ with tensor-product splines (providing explicit roughness control, as in pffr) and estimate them under a generalized spatial two-stage least-squares scheme that instruments the spatially lagged functional response. In this way, our penalization plays the same role as in pffr, stabilizing bivariate functional surfaces with data-driven smoothness, while the estimation targets a different identification problem (spatial autoregressive-type endogeneity) that is not addressed by pffr and related FoFR penalization. Likewise, relative to the P-spline spatial predictor of \cite{Aguilera2017}, our focus is not kriging-style prediction of a functional field but estimation and inference in a regression with an endogenous spatial functional lag; the penalty is integrated into an IV estimator rather than a pure smoothing predictor. Conceptually, PenS2SLS therefore complements existing penalized FoFR approaches by combining, within one estimation routine,    
\begin{inparaenum}
\item[(i)] explicit spline–based smoothness for $\rho$ and $\beta$ with 
\item[(ii)] IV identification for the spatial lag.
\end{inparaenum}

The optimal degree of smoothness in our estimators is governed by smoothing parameters, which are calibrated using an automated grid-search algorithm. This algorithm identifies optimal smoothing parameters by minimizing the Bayesian Information Criterion (BIC), ensuring a data-driven and principled approach to balancing model fit and complexity. By leveraging the BIC, our method effectively selects smoothing parameters that enhance both the accuracy and interpretability of the regression coefficient function estimates, while avoiding issues such as overfitting or under-smoothing. 

To quantify the uncertainty associated with the estimated spatial autoregressive parameter and the bivariate regression coefficient function, we employ a residual-based bootstrap approach. This method constructs confidence intervals by resampling residuals and re-estimating the model parameters, ensuring a data-driven assessment of estimation variability. Our Monte Carlo experiments demonstrate that the bootstrap confidence intervals achieve excellent coverage properties, maintaining high accuracy and stability across various spatial dependence levels. Unlike traditional methods that may suffer from undercoverage or excessive interval width, our approach produces well-calibrated, informative confidence intervals, providing a reliable tool for inference in SFoFR settings. Under some regularity conditions, the $\sqrt{n}$-consistency and asymptotic normality of the proposed estimator are derived. 

The remainder of this paper is structured as follows. In Section~\ref{sec:2}, we introduce the SFoFR model and define its associated parameters. Section~\ref{sec:3} presents a comprehensive discussion of the parameter estimation procedure, including the optimization strategies employed. In Section~\ref{sec:4}, we assess the performance of the proposed approach through an extensive set of Monte Carlo experiments, benchmarking its estimation accuracy and predictive capability against established methods. Section~\ref{sec:5} illustrates the practical applicability of our method through empirical data analysis, with a detailed discussion of the findings. Finally, Section~\ref{sec:6} concludes the article with key insights and avenues for future research.

\section{Model and notations}\label{sec:2}

Let us consider a collection of functional data, denoted as $\{Y_v(t), \X_v(s)\}$, where $v \in \mathcal{D} \subset \mathbb{R}^d$ with $d \geq 1$, observed over a discrete spatial domain $\mathcal{D}$ comprising $n$ spatial locations, indexed by $v_1, \ldots, v_n$. The response $Y_v(t)$ and the predictor $\X_v(s)$ are functional processes that satisfy $Y_v(t) \in \mathcal{L}^p(\mathcal{I}_y)$ and $\X_v(s) \in \mathcal{L}^p(\mathcal{I}_x)$ for all $v$, where $2 \leq p < \infty$. For clarity, we redefine spatial indices using $i$ to denote spatial units $v_i$. Without loss of generality, we assume $\mathcal{I}_y = \mathcal{I}_x = [0,1]$, implying that $Y_v(t)$ and $\X_v(s)$ are real-valued mappings in the unit interval. The observed functional processes are modeled as realizations of mean zero, i.i.d. stochastic processes such that $\mathbb{E}\{Y_v(t)\} = \mathbb{E}\{\X_v(s)\} = 0$. 

To model the spatial functional relationship, we adopt the SFoFR framework as introduced in \citet{BSGARC2024}, as follows:
\begin{equation}\label{eq:sfofrm}
\Y_i(t) = \sum_{j=1}^n w_{ij} \int_0^1 \Y_j(u) \rho(t,u) \, du + \int_0^1 \X_i(s) \beta(t,s) \, ds + \epsilon_i(t),
\end{equation}
where $w_{ij} \in \mathbb{R}^{+}$ represents the $(i,j)$\textsuperscript{th} entry of the $n \times n$ spatial weight matrix $\bm{W} = (w_{ij})_{n \times n}$, characterizing the spatial proximity or influence between locations $i$ and $j$, subject to $w_{ii} = 0$. The function $\rho(t,u) \in \mathcal{C}([0,1]^2)$, where $\mathcal{C}([0,1]^2)$ denotes the space of continuous bivariate functions on $[0,1]^2$, quantifies the spatial autocorrelation structure. Similarly, $\beta(t,s) \in \mathcal{C}([0,1]^2)$ represents the regression coefficient function, encapsulating the influence of the functional predictor $\X_i(s)$ on the response $\Y_i(t)$. The term $\epsilon_i(t)$ signifies a mean-zero random error process that captures unstructured variability. The following remark discusses the completeness of the SFoFR model in~\eqref{eq:sfofrm}.

\begin{remark}\label{rem1}
Under condition $\Vert \rho \Vert_{\infty} < \frac{1}{\Vert \bm{W} \Vert_{\infty}}$, where $\Vert \bm{A} \Vert_\infty$ denotes the maximum absolute row sum of~$\bm{A}$ and $\Vert \rho \Vert_{\infty} := \sup_{(t,u) \in [0,1]^2} \vert \rho(u,t) \vert$, \cite{BSGARC2024} demonstrated that the SFoFR model given in~\eqref{eq:sfofrm} can be equivalently reformulated as:
\begin{equation*}
\Y_i(t) = (\mathbb{I}_d - \mathcal{T})^{-1} \left\{ \int_0^1 \X_i(s) \beta(t,s) \, ds + \epsilon_i(t) \right\},
\end{equation*}
where $\mathcal{T}$ is a bounded linear operator defined as $\mathcal{T}: (\mathcal{L}^p)^n[0,1] \to (\mathcal{L}^p)^n[0,1]$, with its action specified by $(\mathcal{T}\Y)(t) := \bm{W} \int_0^1 \Y(u) \rho(t,u) \, du$. Here, $(\mathcal{L}^p)^n[0,1]$ denotes the space of vector-valued functions $\Y(t) = [\Y_1(t), \ldots, \Y_n(t)]^\top$, where each component $\Y_i(t)$, $i \in \{1, \ldots, n \}$, belongs to $\mathcal{L}^p[0,1]$. $\Vert \Y \Vert_{(\mathcal{L}^p)^n} := \left( \sum_{i=1}^n \Vert \Y_i \Vert_{\mathcal{L}^p}^p \right)^{1/p}$ denote the norm of this space, where $\Vert \Y_i \Vert_{\mathcal{L}^p} := \left( \int_0^1 \vert \Y_i(t) \vert^p \, dt \right)^{1/p}$ is the usual $\mathcal{L}^{p}$ norm. The operator $\mathbb{I}_d$ represents the identity operator in $(\mathcal{L}^p)^n[0,1]$, defined as $\mathbb{I}_d \Y(t) = \Y(t), \forall \Y \in (\mathcal{L}^p)^n[0,1]$. 
\end{remark}

In the formulation of Model~\eqref{eq:sfofrm}, the spatial autocorrelation function $\rho(t,u)$ serves as a pivotal component, encapsulating the intensity and structure of spatial dependence between approximate locations. The specification of spatial weights, $w_{ij} \in \mathbb{R}^+$, for $i, j \in \{1, \ldots, n\}$, is integral to the model, as these weights dictate the spatial connectivity encoded within the matrix $\bm{W} = (w_{ij})_{n \times n}$. The spatial configuration of the $n$ units can exhibit significant heterogeneity, encompassing both regular and irregular arrangements within the spatial domain. Crucially, the construction of $\bm{W}$ must ensure coherence with the underlying spatial topology. 

For spatial domains composed of regularly distributed units, adjacency relationships are typically defined based on shared edges, vertices, or both, as commonly assumed in lattice-based spatial frameworks \citep{Anselin1998}. In contrast, for irregular spatial domains, such as those encountered in polygonal or tessellated structures, adjacency is often established through shared boundaries or other geometric criteria, adhering to definitions prevalent in spatial statistics \citep{Huang2021}.

The elements of the symmetric spatial weight matrix $\bm{W}$, denoted by $w_{ij}$, are frequently constructed using a predefined distance metric, such that $w_{ij} = m(d_{ij})$, where $m(\cdot)$ is a monotonically decreasing function, and $d_{ij}$ quantifies the distance between spatial units $i$ and $j$, satisfying $d_{ii} = 0$. The choice of $d_{ij}$ may reflect a variety of factors, including geographic proximity, economic links, social interactions, or policy-driven relationships, or even a composite measure of them \citep[see, for instance,][]{Yu2016}. Although symmetry in $\bm{W}$ is commonly assumed, it is not a strict requirement; asymmetric weight matrices can be used when spatial structure or data context warrants directional or unbalanced relationships, as outlined in \cite{Huang2021}.

To ensure interpretability and numerical stability, $\bm{W}$ is often row-normalized, giving $w_{ij} = \frac{m(d_{ij})}{\sum_{j=1}^n m(d_{ij})}$, which enforces the constraint that each row sums up to unity. This normalization preserves the inherent sparsity of $\bm{W}$ and ensures that $w_{ii} = 0$, thereby excluding self-contributions in the spatial modeling framework.

\section{Parameter estimation}\label{sec:3}

Our goal is to estimate the unknown bivariate functions $\rho(t,u)$ and $\beta(t,s)$ from discrete and noisy observations. This presents two primary challenges: the infinite-dimensional nature of the functions and the endogeneity introduced by the spatial autoregressive term. We address these challenges using a PenS2SLS procedure, which we detail below.

\subsection{Basis expansion and discretization}

Let $\{\phi_\ell(\cdot)\}_{\ell\ge1}$ and $\{\psi_k(\cdot)\}_{k\ge1}$ denote univariate B-spline bases on $[0,1]$ with fixed orders and quasi-uniform knots. For truncation levels $K_y$ and $K_x$, define the marginal spline spaces
\begin{equation*}
\mathcal{S}_y(K_y) = \mathrm{span}\{\phi_\ell:1\le \ell\le K_y\},\qquad
\mathcal{S}_x(K_x)=\mathrm{span}\{\psi_k:1\le k\le K_x\},
\end{equation*}
and the tensor-product sieve spaces
\begin{equation*}
\mathcal{S}_{\rho}(K_y)=\mathcal{S}_y(K_y)\otimes \mathcal{S}_y(K_y),\qquad
\mathcal{S}_{\beta}(K_y,K_x)=\mathcal{S}_y(K_y)\otimes \mathcal{S}_x(K_x).
\end{equation*}
We approximate the unknown bivariate functions $\rho$ and $\beta$ by their projections onto these sieve spaces:
\begin{equation*}
\rho(t, u) = \sum_{\ell=1}^{K_y} \sum_{m=1}^{K_y} \rho_{\ell m} \phi_{\ell}(t) \phi_m(u), \qquad \beta(t, s) = \sum_{\ell=1}^{K_y} \sum_{k=1}^{K_x} b_{\ell k} \phi_{\ell}(t) \psi_k(s),
\end{equation*}
Consistent with this sieve setup, we estimate the finite-dimensional coefficients $\rho_{\ell m}$ and $b_{\ell k}$, and subsequently recover $\widehat{\rho}(\cdot,\cdot)$ and $\widehat{\beta}(\cdot,\cdot)$ from the basis representations. In line with Assumption $A_2$ in the Appendix, the spline spaces provide controlled approximation error so that $\Vert\rho-\rho_{K}\Vert$ and $\Vert\beta-\beta_{K}\Vert$ vanish at the rates implied by the smoothness of the true functions as $K_y,K_x \to \infty$. Thus, we do not assume the true $\rho(\cdot,\cdot)$ or $\beta(\cdot,\cdot)$ lie exactly in the spline span; instead we work with standard sieve approximations and explicitly account for the (vanishing) approximation error in the asymptotics.

Let the spatially lagged response be denoted by $\widetilde{\Y}_i(u) = \sum_{j=1}^n w_{ij} \Y_j(u)$ for $i \in \{1, \ldots, n\}$. Using this definition, the SFoFR model specified in~\eqref{eq:sfofrm} can be equivalently reformulated as:
\begin{equation}
\Y_{i}(t) = \int_{0}^{1} \widetilde{\Y}_{i}(u) \rho(t, u) \, du + \int_{0}^{1} \X_{i}(s) \beta(t, s) \, ds + \epsilon_{i}(t).\label{eq:sfofrm2}
\end{equation}
We operate under the assumption that the functional data $\Y_i(t)$, $\widetilde{\Y}_i(u)$, and $\X_i(s)$ are densely observed along their respective domains, such that $\Y_i(t)$ corresponds to $\Y_i(t_{i \iota})$, $\widetilde{\Y}_i(u)$ to $\widetilde{\Y}_i(u_{i \iota})$, and $\X_i(s)$ to $\X_i(s_{ir})$, where $\iota = 1, \ldots, M_i$ and $r = 1, \ldots, G_i$. Here, $M_i$ and $G_i$ denote the number of observation points for the curves $\Y_i(t)$, $\widetilde{\Y}_i(u)$, and $\X_i(s)$, respectively. Without loss of generality, we consider the case where $t_{i \iota} = t_{\iota}$, $u_{i \iota} = u_{\iota}$ ($\iota \in \{1, \ldots, M \})$, and $s_{i r} = s_r$ ($r \in \{ 1, \ldots, G \}$).

Let $\Delta_\iota$ and $\Delta_r$ denote the respective lengths of the $\iota$\textsuperscript{th} and $r$\textsuperscript{th} subintervals within the domain $[0,1]$, such that $\Delta_\iota = t_{\iota+1} - t_\iota$ and $\Delta_r = s_{r+1} - s_r$. To approximate the integrals $\int_0^1 \widetilde{\Y}_i(u) \rho(t,u) \, du$ and $\int_0^1 \X_i(s) \beta(t, s) \, ds$ in~\eqref{eq:sfofrm2}, we employ numerical quadrature, resulting in the following discrete formulations:
\begin{center}
\begin{small}
\begin{minipage}{0.45\linewidth}
\begin{align}\label{eq:approx1}
\int_0^1 \widetilde{\Y}_i(u) &\rho(t,u) du \approx \sum_{\iota=1}^{M-1} \bigtriangleup_\iota \rho(t, u_\iota) \widetilde{\Y}_i(u_\iota) \nonumber  \\
&= \sum_{\iota=1}^{M-1} \bigtriangleup_\iota \sum_{\ell=1}^{K_y} \sum_{m=1}^{K_y} \rho_{\ell m} \phi_{\ell}(t) \phi_m(u_\iota) \widetilde{\Y}_i(u_\iota) \nonumber  \\
&= \sum_{\ell=1}^{K_y} \sum_{m=1}^{K_y} \rho_{\ell m} \phi_{\ell}(t) \widetilde{\phi}_{m,i},
\end{align}
\end{minipage}
\qquad
\begin{minipage}{0.45\linewidth}
\begin{align*}
\int_{0}^{1} \X_{i}(s) & \beta(t, s) \, ds \approx \sum_{r=1}^{G-1} \bigtriangleup_{r} \beta(t, s_r) \X_{i}(s_r) \\
&= \sum_{r=1}^{G-1} \bigtriangleup_r \sum_{\ell=1}^{K_y} \sum_{k=1}^{K_x} b_{\ell k} \phi_{\ell}(t) \psi_{k}(s_r) \X_{i}(s_{r}) \\
&= \sum_{\ell=1}^{K_y} \sum_{k=1}^{K_{x}} b_{\ell k} \phi_{\ell}(t) \widetilde{\psi}_{k,i},
\end{align*}
\end{minipage}
\end{small}
\end{center}
\vspace{.1in}
where $\widetilde{\phi}_{m,i} = \sum_{\iota=1}^{M-1} \Delta_\iota \phi_m(u_\iota) \widetilde{\Y}_i(u_\iota)$ and $\widetilde{\psi}_{k,i} = \sum_{r=1}^{G-1} \Delta_r \psi_k(s_r) \X_i(s_r)$. The integrals are thus discretized by expressing the spatial autocorrelation and regression terms as finite sums weighted by the corresponding basis functions $\phi_{\ell}$, $\phi_m$, and $\psi_k$. The coefficients $\rho_{\ell m}$ and $b_{\ell k}$ encode the contributions of the basis expansions in representing the functions $\rho(t,u)$ and $\beta(t,s)$, respectively.

Substituting the approximations in~\eqref{eq:approx1} back into the SFoFR model in~\eqref{eq:sfofrm2} yields a discretized version that is linear in the unknown coefficients:
\begin{align}
\Y_i(t) &\approx \sum_{\ell=1}^{K_y} \sum_{m=1}^{K_y} \rho_{\ell m} \phi_{\ell}(t) \widetilde{\phi}_{m,i} + \sum_{\ell=1}^{K_y} \sum_{k=1}^{K_x} b_{\ell k} \phi_{\ell}(t) \widetilde{\psi}_{k,i} + \epsilon_i(t), \nonumber \\
&= \{ \widetilde{\bm{\phi}}_{i}^\top \otimes \bm{\phi}^\top(t) \} \widetilde{\bm{\rho}} + \{ \widetilde{\bm{\psi}}_{i}^\top \otimes \bm{\phi}^\top(t) \} \widetilde{\bm{b}} + \epsilon_i(t), \label{eq:app2}
\end{align}
where $\bm{\phi}(t) = \{\phi_1(t), \ldots, \phi_{K_y}(t)\}^\top$, $\widetilde{\bm{\phi}}_{i} = (\widetilde{\phi}_{1,i}, \ldots, \widetilde{\phi}_{K_y, i})^\top$, $\widetilde{\bm{\psi}}_{i} = (\widetilde{\psi}_{1,i}, \ldots, \widetilde{\psi}_{K_x, i})^\top$, $\widetilde{\bm{\rho}} = \text{vec}(\bm{\rho})$, with $\bm{\rho}$ being a $K_y \times K_y$ matrix whose the $(\ell, m)$\textsuperscript{th} element is $\rho_{\ell m}$, $\ell, m \in \{1, \ldots, K_y\}$, $\widetilde{\bm{b}} = \text{vec}(\bm{b})$, with $\bm{b}$ being a $K_y \times K_x$ matrix whose $(\ell, k)$\textsuperscript{th} element is $b_{\ell k}$, $\ell \in \{1, \ldots, K_y\}$ and $k \in \{1, \ldots, K_x\}$, and $\otimes$ denotes the Kronecker product. 

\subsection{The penalized spatial two-stage least-squares estimator}

The presence of the spatially lagged response term $\widetilde{\Y}_i(u)$ in~\eqref{eq:sfofrm2} introduces an endogeneity problem. Because $\widetilde{\Y}_i(u)$ is a weighted average of neighboring observations $\Y_j(u)$, which themselves depend on the error terms $\epsilon_j(t)$, the regressor associated with $\rho(t,u)$ in~\eqref{eq:app2} is correlated with the model error $\epsilon_i(t)$. A standard least-squares estimation would therefore yield biased and inconsistent estimates.

To address this, we adopt the generalized spatial two-stage least-squares framework and extend it to our functional setting. The core idea is to use IVs that are correlated with the endogenous regressor but uncorrelated with the error term.

In the first stage, we construct valid IVs and project the endogenous regressor onto the space spanned by them. Natural candidates for IVs are the spatially lagged versions of the exogenous functional predictor as they are correlated with the spatial structure of the model but are, by assumption, uncorrelated with the error term $\epsilon(t)$. In the approximate model~\eqref{eq:app2}, the first term on the right-hand side, that is, $\widetilde{\bm{\phi}}_{i}^\top \otimes \bm{\phi}^\top(t)$, is the endogenous variable, while the second term, $\widetilde{\bm{\psi}}_{i}^\top \otimes \bm{\phi}^\top(t)$, is the exogenous variable. We construct the matrix of basis expansion coefficients $\bm{\Pi}$ as follows:
\begin{equation*}
\bm{\Pi} =
\begin{pmatrix}
\widetilde{\bm{\psi}}_1 \otimes \bm{\phi}^* & \widetilde{\bm{\phi}}_1 \otimes \bm{\phi}^* \\
\widetilde{\bm{\psi}}_2 \otimes \bm{\phi}^* & \widetilde{\bm{\phi}}_2 \otimes \bm{\phi}^* \\
\vdots & \vdots \\
\widetilde{\bm{\psi}}_n \otimes \bm{\phi}^* & \widetilde{\bm{\phi}}_n \otimes \bm{\phi}^*
\end{pmatrix},
\end{equation*}
where $\bm{\phi}^*$ is an $M \times K_y$ matrix comprising $\{\bm{\phi}(t_1), \ldots, \bm{\phi}(t_M)\}$ as its rows. The basis expansion coefficients are then obtained by projecting $\bm{\Pi}$ onto the space spanned by a set of instrumental variables IVs, which typically include $\{\widetilde{\bm{\psi}}_{i}^\top \otimes \bm{\phi}^\top(t)\}$ and discretized versions of the spatially lagged functional predictor.

Let $\widetilde{\X}^{(q)}(s) = \bm{W}^{q} \X(s)$, $q \in \{1, \ldots, Q\}$, denote the $q$\textsuperscript{th} order spatially lagged functional predictor, where $Q$ is the number of IVs to be considered in the model. Furthermore, let $\bm{Z} = (\bm{Z}_0, \bm{Z}_1, \ldots, \bm{Z}_Q)^\top$, where $\bm{Z}_0 = \{ \widetilde{\bm{\psi}}_{1}^\top \otimes \bm{\phi}^\top(t), \ldots, \widetilde{\bm{\psi}}_{n}^\top \otimes \bm{\phi}^\top(t) \}^\top$, and $\bm{Z}_q = (\widetilde{\bm{\psi}}^{(q)})^\top \otimes \bm{\phi}^\top(t)$ with $\widetilde{\bm{\psi}}^{(q)} = (\widetilde{\bm{\psi}}_1^{(q)}, \ldots, \widetilde{\bm{\psi}}_n^{(q)})^\top$, the $i$\textsuperscript{th} element of which is
\begin{align*}
\widetilde{\bm{\psi}}_{i}^{(q)} &= (\widetilde{\psi}_{1,i}^{(q)}, \ldots, \widetilde{\psi}_{K_x, i}^{(q)})^\top \\ 
\widetilde{\psi}_{k,i}^{(q)} &= \sum_{r=1}^{G-1} \Delta_r \psi_k(s_r) \widetilde{\X}_i^{(q)}(s_r),
\end{align*}
for $q \in \{1, \ldots, Q\}$, $k \in \{1, \ldots, K_x\}$, and $i \in \{1, \ldots, n\}$. We then project $\bm{\Pi}$ onto the space spanned by the instruments $\bm{Z}$ to obtain the fitted or "purged" design matrix, $\widehat{\bm{\Pi}}$:
\begin{equation*}
\widehat{\bm{\Pi}} = \bm{Z} (\bm{Z}^\top \bm{Z})^{-1} \bm{Z}^\top \bm{\Pi}.
\end{equation*}
The matrix $\widehat{\bm{\Pi}}$ now contains regressors that are, by construction, uncorrelated with the error term.

In the second stage, we estimate the parameter vector $\bm{\theta} = (\widetilde{\bm{\rho}}^\top, \widetilde{\bm{b}}^\top)^\top$ by minimizing a penalized least-squares objective function. To prevent overfitting and ensure the estimated functions $\widehat{\rho}(t,u)$ and $\widehat{\beta}(t,s)$ are smooth, we add a roughness penalty to the objective function. We write $\bm{\Phi}_t = \int_0^1 \bm{\phi}(t) \bm{\phi}(t)^\top dt$, $\bm{\Phi}_u = \bm{\Phi}_t$, $\bm{\Psi} = \int_0^1 \bm{\psi}(s) \bm{\psi}(s)^\top ds$, and let $\bm{D}_t,\bm{D}_u,\bm{D}_s$ be the usual marginal derivative-penalty matrices; the resulting tensor-product penalty takes the standard Kronecker-sum form used for anisotropic smoothers:
\begin{align*}
(\bm{D}_u)_{m, \widetilde{m}} &= \int_0^1 \phi_m^{''}(u) \phi_{\widetilde{m}}^{''}(u) \, du, \qquad m, \widetilde{m} \in \{1, \ldots, K_y\}, \\
(\bm{D}_t)_{\ell, \widetilde{\ell}} &= \int_0^1 \phi_{\ell}^{''}(t) \phi_{\widetilde{\ell}}^{''}(t) \, dt, \qquad \ell, \widetilde{\ell} \in \{1, \ldots, K_y\}, \\
(\bm{D}_s)_{k, \widetilde{k}} &= \int_0^1 \psi_k^{''}(s) \psi_{\widetilde{k}}^{''}(s) \, ds, \qquad k, \widetilde{k} \in \{1, \ldots, K_x\},
\end{align*}
where $^{''}$ denotes the second-order derivative. Let us define the penalties for the bivariate coefficient functions $\rho(t,u)$ and $\beta(t,s)$, denoted by $\bm{P}(\rho)$ and $\bm{P}(\beta)$, respectively, based on the integrated squared second derivatives of the functions as follows:
\begin{align*}
\bm{P}(\rho) &= \int_0^1 \int_0^1 \left\lbrace \frac{\partial^2}{\partial t^2} \rho(t,u) \right\rbrace^2 du \, dt 
+ \int_0^1 \int_0^1 \left\lbrace \frac{\partial^2}{\partial u^2} \rho(t,u) \right\rbrace^2 du \, dt 
= \widetilde{\bm{\rho}}^\top \left( \bm{\Phi}_u \otimes \bm{D}_t + \bm{D}_u \otimes \bm{\Phi}_t \right) \widetilde{\bm{\rho}}, \\
\bm{P}(\beta) &= \int_0^1 \int_0^1 \left\lbrace \frac{\partial^2}{\partial t^2} \beta(t,s) \right\rbrace^2 ds \, dt 
+ \int_0^1 \int_0^1 \left\lbrace \frac{\partial^2}{\partial s^2} \beta(t,s) \right\rbrace^2 ds \, dt 
= \widetilde{\bm{b}}^\top \left( \bm{\Psi} \otimes \bm{D}_t + \bm{D}_s \otimes \bm{\Phi}_t \right) \widetilde{\bm{b}}.
\end{align*}

The second-stage objective function to estimate $\widetilde{\bm{\rho}}$ and $\widetilde{\bm{b}}$ is as follows:
\begin{align*}
\mathcal{G}(\widetilde{\bm{\rho}}, \widetilde{\bm{b}}) &= \Vert \bm{Z}^* \{\text{vec}(\widetilde{\Y}) - \bm{\Pi} \bm{\theta} \} \Vert^2 + \frac{1}{2} \lambda_{\rho} \bm{P}(\rho) + \frac{1}{2} \lambda_{\beta} \bm{P}(\beta), \\
&= \{\text{vec}(\widetilde{\Y}) - \bm{\Pi} \bm{\theta} \}^\top \bm{Z}^* \{\text{vec}(\widetilde{\Y}) - \bm{\Pi} \bm{\theta} \} + \frac{1}{2} \lambda_{\rho} \bm{P}(\rho) + \frac{1}{2} \lambda_{\beta} \bm{P}(\beta),
\end{align*}
where $\bm{Z}^* = \bm{Z} (\bm{Z}^\top \bm{Z})^{-1} \bm{Z}^\top$, and $\widetilde{\Y}$ is constructed as follows:
\begin{equation*}
\widetilde{\Y} = 
\begin{pmatrix}
\Y_1(t_1) & \Y_2(t_1) & \dots & \Y_n(t_1) \\
\Y_1(t_2) & \Y_2(t_2) & \dots & \Y_n(t_2) \\
\vdots & \vdots & \ddots & \vdots \\
\Y_1(t_M) & \Y_2(t_M) & \dots & \Y_n(t_M)
\end{pmatrix}.
\end{equation*}
Simplifying the objective function, we obtain:
\begin{equation*}
\mathcal{G}(\widetilde{\bm{\rho}}, \widetilde{\bm{b}}) = \text{vec}(\widetilde{\Y})^\top \bm{Z}^* \text{vec}(\widetilde{\Y}) - 2 \bm{\theta}^\top \underset{\widehat{\bm{\Pi}}^\top}{\underbrace{\bm{\Pi}^\top \bm{Z}^*}} \text{vec}(\widetilde{\Y}) + \bm{\theta}^\top \underset{\widehat{\bm{\Pi}}^\top}{\underbrace{\bm{\Pi}^\top \bm{Z}^*}} \bm{\Pi} \bm{\theta} + \frac{1}{2} \lambda_{\rho} \bm{P}(\rho) + \frac{1}{2} \lambda_{\beta} \bm{P}(\beta).
\end{equation*}

Taking the derivative of $\mathcal{G}(\widetilde{\bm{\rho}}, \widetilde{\bm{b}})$ with respect to $\bm{\theta} = (\widetilde{\bm{\rho}}^\top, \widetilde{\bm{b}}^\top)^\top$ and setting it to zero yields the solution:
\begin{equation*}
\frac{\partial \mathcal{G}(\widetilde{\bm{\rho}}, \widetilde{\bm{b}})}{\partial \bm{\theta}} = -2 \underset{\widehat{\bm{\Pi}}^\top}{\underbrace{\bm{\Pi}^\top \bm{Z}^*}} \text{vec}(\widetilde{\Y}) + 2 \underset{\widehat{\bm{\Pi}}^\top}{\underbrace{\bm{\Pi}^\top \bm{Z}^*}} \bm{\Pi} \bm{\theta} + \bm{R}(\lambda_{\rho}, \lambda_{\beta}) \bm{\theta} = 0,
\end{equation*}
where $\bm{R}(\lambda_{\rho}, \lambda_{\beta})$ is the block diagonal penalty matrix given by:
\begin{equation*}
\bm{R}(\lambda_{\rho}, \lambda_{\beta}) = 
\begin{pmatrix}
\lambda_{\rho} (\bm{\Phi}_u \otimes \bm{D}_t + \bm{D}_u \otimes \bm{\Phi}_t) & 0 \\
0 & \lambda_{\beta} (\bm{\Psi} \otimes \bm{D}_t + \bm{D}_s \otimes \bm{\Phi}_t)
\end{pmatrix}.
\end{equation*}

Because the projection matrix is idempotent, we obtain the linear system as follows:
\begin{equation*}
\Big\{ \underset{\widehat{\bm{\Pi}}^\top}{\underbrace{\bm{\Pi}^\top \bm{Z}^*}} \bm{\Pi} + \bm{R}(\lambda_{\rho}, \lambda_{\beta}) \Big\} \bm{\theta} = \underset{\widehat{\bm{\Pi}}^\top}{\underbrace{\bm{\Pi}^\top \bm{Z}^*}} \text{vec}(\widetilde{\Y}).
\end{equation*}
The PenS2SLS estimator for $\bm{\theta}$ is therefore:
\begin{equation*}
\widehat{\bm{\theta}} = \Big\{ \underset{\widehat{\bm{\Pi}}^\top}{\underbrace{\bm{\Pi}^\top \bm{Z}^*}} \bm{\Pi} + \bm{R}(\lambda_{\rho}, \lambda_{\beta}) \Big\}^{-1} \underset{\widehat{\bm{\Pi}}^\top}{\underbrace{\bm{\Pi}^\top \bm{Z}^*}} \text{vec}(\widetilde{\Y}).
\end{equation*}
Finally, the estimated coefficient functions are reconstructed by plugging the estimated coefficients $\widehat{\widetilde{\bm{\rho}}}$ and $\widehat{\widetilde{\bm{b}}}$ back into their respective basis expansions, as follows:
\begin{align*}
\widehat{\rho}(t,u) &= \{\bm{\phi}^\top(u) \otimes \bm{\phi}^\top(t)\} \widehat{\widetilde{\bm{\rho}}} \\
\widehat{\beta}(t,s) &= \{\bm{\psi}^\top(s) \otimes \bm{\phi}^\top(t)\} \widehat{\widetilde{\bm{b}}},
\end{align*}
where $\bm{\phi}(u) = \{\phi_1(u), \ldots, \phi_{K_y}(u)\}^\top$ and $\bm{\psi}(s) = \{\psi_1(s), \ldots, \psi_{K_x}(s)\}^\top$.

We emphasize that the tensor-product B-spline representation $\beta(t,s) = \sum_{\ell=1}^{K_t}\sum_{k=1}^{K_s} b_{\ell k} \phi_{\ell}(t) \psi_k(s)$ does not impose a separable form $\beta(t,s) = \beta_1(t) \beta_2(s)$. Separability would require $\{b_{\ell k}\}$ to be rank–1, whereas our estimator admits a full matrix of coefficients and therefore arbitrary smooth interactions between $t$ and $s$. This is the standard way tensor-product smooths represent bivariate interactions without scale-arbitration issues, cf. low-rank, scale-invariant tensor products in generalized additive models \citep[see, e.g.,][]{Wood2003, Wood2006}. 

Consistent with common practice for tensor-product smooths, our roughness penalty is the anisotropic Kronecker-sum of marginal penalties,
\begin{equation*}
\bm{P}(\beta) = \lambda_t \big\| \partial^{m_t}_t \beta\big\|^2 + \lambda_s \big\| \partial^{m_s}_s \beta\big\|^2 \equiv \lambda_t (\bm{\Psi} \otimes \bm{D}_t) + \lambda_s (\bm{D}_s \otimes \bm{\Phi}_t),
\end{equation*}
which controls wiggliness along each axis and still allows fully nonseparable $\beta$; no mixed derivative term is required for interactions to be represented. This construction is widely used because it is invariant to the units of $t$ and $s$, numerically stable, and interpretable (one smoothing parameter per margin) \citep{Wood2006}. An alternative is an isotropic thin-plate penalty that includes mixed partials; both approaches are well established, but anisotropic tensor penalties are typically preferred when the two axes differ in scale or smoothness \citep{Wood2003}. Our simulations deliberately use nonseparable truth to demonstrate that the proposed estimator recovers interacting surfaces under this penalty structure, consistent with theory for penalized FoFR models and their identifiability conditions \citep{ivanescu2015}. The same remarks apply to the spatial kernel $\rho(t,u)$, which is modeled with a full tensor-product basis (not constrained to be separable) and the same anisotropic penalty.

The following theorem discusses the consistency and asymptotic normality properties of the proposed estimators $\widehat{\rho}(u,t)$ and $\widehat{\beta}(s,t)$. The proof of Theorem~\ref{th:1} is outlined in the Appendix.
\begin{theorem}\label{th:1}
Let $\widehat{\rho}(u,t)$ and $\widehat{\beta}(s,t)$ denote the Pen2SLS estimates of the true parameter functions $\rho(t,u)$ and $\beta(t,s)$, respectively. Under the assumptions $A_1$-$A_5$ given in the Appendix, $\widehat{\rho}(u,t)$ and $\widehat{\beta}(s,t)$ are $\sqrt{n}$-consistent and asymptotically normal:
\begin{align*}
\sqrt{n} \{\widehat{\rho}(t,u) - \rho(t,u) \} \xrightarrow{d} \mathcal{GP} \{0, \bm{\Sigma}_\rho (t,u; t^\prime, u^\prime) \}, \\
\sqrt{n} \{\widehat{\beta}(t,s) - \beta(t,u) \} \xrightarrow{d} \mathcal{GP} \{0, \bm{\Sigma}_\beta (t,s; t^\prime, s^\prime) \},
\end{align*}
where the covariance kernels of the asymptotic Gaussian process ($\mathcal{GP}$), $\bm{\Sigma}_\rho (t,u; t^\prime, u^\prime)$, and $\bm{\Sigma}_\beta (t,s; t^\prime, s^\prime)$, are defined in the Appendix.
\end{theorem}

\begin{remark}
It is important to interpret the result of Theorem~\ref{th:1} with caution. Our asymptotic statements distinguish between (a) the infinite-dimensional coefficient surface $\beta(t,s)$ and $\rho(t,u)$ viewed as functions and (b) finite-dimensional linear functionals of these objects. In line with standard nonparametric theory, the full functions are estimated at rates determined by their smoothness, which are generally slower than $\sqrt{n}$ \citep[see, for example, the canonical results for nonparametric regression and for functional linear regression slope estimation, see][]{Hall2007}. By contrast, when the target is a smooth linear functional of the form $L(\beta)=\iint \omega(t,s) \beta(t,s) dt ds$ or $L(\rho)=\iint v(t,u) \rho(t,u) dt du$, penalized series/sieve procedures admit $\sqrt{n}$-consistent, asymptotically normal inference under standard regularity and tuning conditions \citep[see, e.g.,][]{Chen2007}. Accordingly, throughout we reserve the $\sqrt{n}$ claim only for such linear functionals (and for fixed-$K$ coefficient vectors), while convergence of the entire surfaces is stated in appropriate $L_2$-type norms at the usual nonparametric rates.
\end{remark}

\subsection{Selection of instrumental variables}\label{sec:3.1}

The selection of IVs is crucial to address the endogeneity introduced by the spatially lagged functional response term, $\int_0^1 \widetilde{\Y}_i(u) \rho(t, u) \, du$. In the SFoFR model, the endogenous variable, $ \widetilde{\bm{\phi}}_{i}^\top \otimes \bm{\phi}^\top(t)$, is intrinsically linked to the error term $\epsilon_i(t)$ through the spatial autocorrelation structure. To address this, we propose the construction of IVs from spatially lagged functional predictors and responses, taking into account their relevance and exogeneity.

The spatially lagged functional predictors, 
$\widetilde{\X}_i^{(q)} = \bm{W}^q \X_{i}(s)$, serve as natural IV candidates, with their basis coefficients: 
\begin{equation*}
\widetilde{\psi}_{k,i}^{(q)} = \sum_{r=1}^{G-1} \Delta_r \psi_k(s_r) \widetilde{\X}_i^{(q)}(s_r), \qquad k \in \{1, \ldots, K_x \}, 
\end{equation*}
capturing both the functional and spatial dependencies in $\X_i(s)$. These lagged variables are relevant due to their recursive dependence on the spatial structure, but must be chosen judiciously to avoid weak instrument issues.

While it is theoretically possible to construct many IVs arbitrarily by increasing the order of spatial lags $(q > 1)$, higher-order lags tend to weaken the instruments, as the correlation between $\widetilde{\X}_i^{(q)}(s)$ and $ \widetilde{\bm{\phi}}_{i}^\top \otimes \bm{\phi}^\top(t) $ diminishes with $q$. This trade-off requires careful selection of $q$, balancing instrument strength and the need for sufficient IVs to consistently estimate coefficient functions $\rho(t,u)$ and $\beta(t,s)$. In practice, $q$ should remain low to ensure IV relevance, particularly when the spatial autocorrelation structure of the data decays rapidly.

As $n$ increases, the theoretical requirements for $K_x$ and $K_y$ also evolve. To consistently estimate the smooth coefficient functions $\rho(t,u)$ and $\beta(t,s)$, the truncation parameters $K_x$ and $K_y$ must increase with $n$ to capture the underlying functional relationships (see assumption $A_3~(i))$. The dimensionality of IVs, determined by $K_x$, $K_y$, and $Q$, must align with practical sample sizes. For most empirical settings, where $n$ is moderate and $K_x, K_y < 10$ sufficient IVs can be constructed without compromising the validity of the model \citep[see, e.g.,][for more discussion about the selection of IVs]{Hoshino2024}.

\subsection{Selection of smoothing and truncation parameters using BIC}\label{sec:3.2}

The performance of the PenS2SLS estimator in the SFoFR model relies on the appropriate choice of the smoothing parameters ($\lambda_{\rho}$, $\lambda_{\beta}$) that govern the smoothness of the estimated coefficient functions. Suboptimal parameter selection can lead to overfitting or underfitting, adversely affecting the accuracy and interpretability of estimated functions $\rho(t,u)$ and $\beta(t,s)$. Therefore, a robust and data-driven method for parameter selection is crucial to ensure optimal model performance.

To address this need, we adopt the BIC, computed on the basis of the log-likelihood, expressed via the squared error terms. The BIC criterion evaluates the trade-off between model fit and complexity by penalizing excessive flexibility while avoiding overfitting. By minimizing the BIC, we can systematically determine the optimal values of $\lambda_{\rho}$ and $\lambda_{\beta}$ that produce a balance between the fidelity to observed data and the smoothness of the estimated coefficient functions.

Let us respectively denote by $\bm{Y}_i$ and $\bm{\epsilon}_i$ the values of $\Y_i(t)$ and $\epsilon_i(t)$ recorded at discrete time points $t \in \{t_1, \ldots, t_M \}$ for $i \in \{1, \ldots, n\}$. For simplicity, $\bm{\epsilon}_i$ is assumed to be sampled from $\mathcal{N}(0, \sigma^2 \mathbb{I})$, where $\mathbb{I}$ denotes the identity matrix. Then, for a model fitted by the smoothing parameters ($\lambda_{\rho}$, $\lambda_{\beta}$), the likelihood function for the $i$\textsuperscript{th} sample is defined by
\begin{equation}
f(\bm{Y}_i \vert \bm{\Theta}_{\lambda_{\rho}, \lambda_{\beta}}) = \frac{1}{(2 \pi \sigma_{\lambda_{\rho}, \lambda_{\beta}})^{n/2}} \exp \Big\lbrace - \frac{(\bm{Y}_i - \widehat{\bm{Y}}_i )^\top (\bm{Y}_i - \widehat{\bm{Y}}_i )}{2 \sigma_{\lambda_{\rho}, \lambda_{\beta}}^2} \Big\rbrace, \label{eq:lik}
\end{equation}
where $\bm{\Theta}_{\lambda_{\rho}, \lambda_{\beta}}$ is the vector of the model parameters and $\widehat{\bm{Y}}_i$ is the prediction of $\bm{Y}_i$, that is, the values of $\widehat{\Y}_i(t)$ calculated as follows:
\begin{equation*}
\widehat{\Y}_i(t) = (\mathbb{I}_d - \mathcal{T})^{-1} \left\{ \int_0^1 \X_i(s) \widehat{\beta}_{\lambda_{\rho}, \lambda_{\beta}}(t,s) \, ds \right\},
\end{equation*}
where $\widehat{\beta}_{\lambda_{\rho}, \lambda_{\beta}}(t,s)$ is the estimated parameter function obtained by the smoothing parameters ($\lambda_{\rho}$, $\lambda_{\beta}$). Define the $\log$-likelihood for~\eqref{eq:lik} by $\mathcal{L}(\bm{\Theta}_{\lambda_{\rho}, \lambda_{\beta}}) = \sum_{i=1}^n \log \{f(\bm{Y}_i \vert \bm{\Theta}_{\lambda_{\rho}, \lambda_{\beta}}) \}$. Then, the BIC for a given $\lambda_{\rho}$ and $\lambda_{\beta}$ is defined as follows:
\begin{equation}\label{eq:bic}
\text{BIC} = -2 \mathcal{L}(\bm{\Theta}_{\lambda_{\rho}, \lambda_{\beta}}) + \omega \log(n),
\end{equation}
where $\omega$ is the degree of freedom of the model, which depends on the smoothing parameters.

A question naturally arises about the choice of the PenS2SLS estimator for parameter estimation while using a likelihood-based BIC for selecting the smoothing parameters. Our estimator targets the spatial endogeneity in~\eqref{eq:sfofrm2} via the PenS2SLS criterion, rather than maximum likelihood, for two practical reasons: 
\begin{inparaenum}
\item[(i)] generalized spatial two-stage least-squares procedures are consistent and computationally simple for spatial autoregressive structure without requiring full distributional specification \citep{Kelejian1998}; and \item[(ii)] a full likelihood for the functional spatial autoregressive model with bivariate spline surfaces $(\rho,\beta)$ is high-dimensional and numerically intensive. 
\end{inparaenum}

For smoothing-parameter selection we adopt a Gaussian quasi-likelihood view of the conditional mean fitted by PenS2SLS and evaluate a BIC of the form in~\eqref{eq:bic}, where $\mathcal{L}(\bm{\Theta}_{\lambda_{\rho}, \lambda_{\beta}})$ is the Gaussian quasi-log-likelihood computed from PenS2SLS residuals. This mirrors standard practice in penalized spline smoothing, where likelihood (or quasi-likelihood) criteria are used to tune penalties even when estimation is not reliant on a pure maximum likelihood estimator \citep{Wood2016}. Conceptually, the BIC here acts as a quasi–BIC; it scores the same fitted mean that PenS2SLS produces, while the penalty term accounts for model complexity through effective degrees of freedom rather than raw parameter counts.

An alternative would be to use information criteria tailored to two stage least-squares (or generalized method of moment)-based on the $J$ statistic (e.g., the BIC analogue of \citealp{Andrew2001}), which we view as complementary. We prefer quasi-BIC for $\lambda$-selection because 
\begin{inparaenum}
\item[(i)] it directly balances fit against smoothness for the penalized spline surfaces $(\rho,\beta)$, and 
\item[(ii)] it aligns with widespread smoothing parameter selection practice for penalized regression in non-Gaussian and dependent–data settings \citep{Wood2016}.
\end{inparaenum}

\section{Monte Carlo experiments}\label{sec:4}

We perform an extensive series of Monte Carlo simulations to assess the estimation accuracy and the predictive performance of the proposed SFoFR method. The finite-sample properties of our approach are benchmarked against two alternative methodologies: 
\begin{inparaenum}
\item[(i)] the SFoFR approach introduced by \cite{BSGARC2024}; and 
\item[(ii)] the widely utilized pffr method proposed by \cite{ivanescu2015} and \cite{refund}. 
\end{inparaenum}
It is important to note that, unlike our proposed method and the SFoFR approach by \cite{BSGARC2024}, the pffr method does not account for spatial dependencies within the model framework, thus offering a comparative perspective on the importance of spatial dependence in functional regression models. Computationally, the implementation of the proposed method is documented in the \texttt{robflreg} package \citep{BS25} in \Rlogo.

In our Monte Carlo experiments, we adopt the data generation process described in \cite{BSGARC2024}, where both the functional predictor $\X_i(s)$ and the functional response$ \Y_i(t)$ are evaluated at 101 equally spaced points over the interval $[0,1]$, with $s, t = \frac{r}{101}$, where $r \in \{1, \ldots, 101\}$. The functional predictor $\X_i(s)$ is constructed using a truncated Fourier series representation as follows:
\begin{equation*}
\X_i(s) = \sum_{k=1}^{10} \frac{1}{k^{3/2}} \left\{ \nu_{i1,k} \sqrt{2} \cos(k \pi s) + \nu_{i2,k} \sqrt{2} \sin(k \pi s) \right\},
\end{equation*}
where $\nu_{i1,k}$ and $\nu_{i2,k}$ are independent standard normal random variables, ensuring variability between realizations. The bivariate regression coefficient function $\beta(t,s)$ is specified as 
\begin{equation*}
\beta(t,s) = 2 + s + t + 0.5 \sin(2 \pi s t).
\end{equation*}

The spatial autocorrelation function $\rho(t,u)$ is defined as 
\begin{equation*}
\rho(t,u) = \eta \frac{1 + u t}{1 + \vert u - t \vert}, 
\end{equation*}
where $\eta \in (0,1)$ regulates the strength of spatial dependence. A lower value of $\eta$ (e.g., $\eta$ =~0.1) corresponds to a weak spatial correlation, indicating minimal interaction between observations. In contrast, a larger value of $\eta$ (e.g., $\eta = 0.9$) reflects a pronounced spatial dependence, signifying a strong influence of neighboring observations. To evaluate the impact of varying spatial effects, we consider three scenarios with weak, moderate, and strong spatial dependencies, corresponding to $\eta \in \{0.1, 0.5, 0.9\}$.

To generate the spatial weight matrix $\bm{W}$, we employ the inverse distance matrix, where each entry $w_{i j}$ represents the inverse of the distance between locations $i$ and $j$. Specifically, for $i \neq j$, the weights are defined as $w_{i j} = \frac{1}{1 + \vert i - j \vert}$. The diagonal elements $w_{i i}$ are set to zero for all $i$, ensuring that there is no self-weighting. To maintain consistency and interpretability, the matrix $\bm{W}$ is row-normalized so that the weights of each location sum up to one. This is achieved using: $\bm{W}_{i.} = \frac{w_{ij}}{\sum_{j=1}^{n} w_{ij}}, \forall i.$ The resulting matrix $\bm{W}$ ensures proper scaling of spatial influences while preserving the relative spatial relationships across all locations.

The functional response $\Y_{i}(t)$ is generated using the Neumann series approximation, as outlined in \cite{Hoshino2024}, utilizing the contraction property of the spatial autocorrelation function $\rho(t,u)$ established in Remark~\ref{rem1}. Specifically, the response is defined as 
\begin{equation*}
\Y_i(t) = (\mathbb{I}_d - \mathcal{T})^{-1} \Big\{\int_0^1 \X_i(s) \beta(t,s) \, ds + \epsilon_i(t)\Big\}, 
\end{equation*}
with $\epsilon_{i}(t)$ representing independent standard normal noise. The operator $\mathcal{T}$ is determined by the spatial weight matrix $\bm{W}$, and the contraction condition $\Vert \rho \Vert_{\infty} < 1 / \Vert \bm{W} \Vert_{\infty} $ ensures that the Neumann series converges to a unique solution. To approximate the series, the functional response is computed iteratively as 
\begin{equation*}
\Y_i^{(\mathcal{M})}(t) = \sum_{m=0}^\mathcal{M} \mathcal{T}^m \Big\{\int_0^1 \X_i(s) \beta(t,s) \, ds + \epsilon_i(t)\Big\}, 
\end{equation*}
where $\mathcal{M}$ is incremented until the convergence criterion, $\max_{i \in \{1, \ldots, n\}} \big\vert \Y_i^{(\mathcal{M})}(t) - \Y_i^{(\mathcal{M}-1)}(t) \big\vert < 0.001$, is satisfied for all $t$. This iterative process ensures that the functional response generated captures the desired spatial correlation structure while maintaining numerical stability. By truncating the Neumann series at an appropriate level of $\mathcal{M}$, the method efficiently approximates the spatially dependent functional response.

In Monte Carlo experiments, we evaluated the performance of the proposed method under three different training sample sizes: $n_{\text{train}} \in \{100, 250, 500\}$. For each training dataset, the models are constructed and their estimation accuracy is assessed. Specifically, we quantify the estimation performance of the bivariate regression coefficient function $\beta(t,s)$ and the spatial autocorrelation function $\rho(t,u)$ using the root relative integrated squared percentage estimation error (RRISPEE), defined as follows:
\begin{align*}
\text{RRISPEE}(\widehat{\beta}) &= 100 \times \sqrt{\frac{\Vert \beta(t,s) - \widehat{\beta}(t,s) \Vert_2^2}{\Vert \beta(t,s) \Vert_2^2}}, \\ 
\text{RRISPEE}(\widehat{\rho}) &= 100 \times \sqrt{\frac{\Vert \rho(t,u) - \widehat{\rho}(t,u) \Vert_2^2}{\Vert \rho(t,u) \Vert_2^2}}, 
\end{align*}
where $\Vert \cdot \Vert_2$ denotes the $\mathcal{L}^2$-norm. The RRISPEE measures the percentage error relative to the true functions, providing a standardized and interpretable metric to compare the accuracy of the estimates in different sample sizes.

For each training sample size, $n_{\text{test}} = 1000$ independent samples are generated to form the test dataset. The predictive performance of the methods is then evaluated by applying the fitted models, trained on the respective training datasets, to these test samples. The predictive accuracy is quantified using the root mean squared percentage error (RMSPE), defined as follows:
\begin{equation*}
\text{RMSPE} = 100 \times \sqrt{\frac{\Vert \Y(t) - \widehat{\Y}^{\text{new}}(t) \Vert_2^2}{\Vert \Y(t) \Vert_2^2}},
\end{equation*}
where $\Y(t)$ denotes the true functional response in the test data set, and $\widehat{\Y}^{\text{new}}(t)$ represents the predicted response obtained from the fitted models. The training and test data sets are strictly mutually exclusive, ensuring unbiased out-of-sample predictions. In scenarios where there is overlap between the training and test data sets, a trend-corrected approach, such as the method proposed by \cite{Goulard2017}, can mitigate potential bias and maintain the validity of out-of-sample predictions.

To further investigate the performance of our proposed PSFoFR model, we employ a residual-based bootstrap approach to assess the uncertainty associated with the estimated bivariate regression coefficient function $\beta(t,s)$ and the spatial autocorrelation parameter function $\rho(t,u)$. Specifically, bootstrap confidence intervals (CIs) are constructed to quantify the estimation accuracy of these parameters. The procedure involves resampling residuals from the fitted model. Let the residuals of the initial model fit be
$\widehat{\epsilon}_i(t) = \Y_i(t) -\widehat{\Y}_i(t)$. For each bootstrap iteration $\kappa$, (centered) residuals $ \widehat{\epsilon}_i^*(t)=\widehat{\epsilon}_i(t)-\frac{1}{n}\sum^n_{i=1}\widehat{\epsilon}_i(t)$ are sampled with replacement to generate bootstrap responses, 
\begin{equation*}
\widehat{\Y}_i^*(t) = (\mathbb{I}_d - \mathcal{T})^{-1} \left\{ \int_0^1 \X_i(s) \widehat{\beta}_{\lambda_{\rho}, \lambda_{\beta}}(t,s) \, ds \right\} + \widehat{\epsilon}_i^*(t). 
\end{equation*}
The model is re-estimated using the bootstrap responses to obtain a collection of bootstrap estimates $\{\widehat{\beta}^*_\kappa(t,s),~ \widehat{\rho}^*_\kappa(t,u) \}_{\kappa = 1}^B$, where $B$ is the number of bootstrap samples. For each $(t,s)$ and $(t,u)$, the bootstrap confidence intervals at the nominal level $(1-\alpha)$ are constructed as follows:
\begin{align*}
\text{CI}_\beta(t,s) &= \left[\widehat{\beta}^*_{\alpha/2}(t,s), \quad \widehat{\beta}^*_{1-\alpha/2}(t,s) \right] \\ 
\text{CI}_\rho(t,u) &= \left[\widehat{\rho}^*_{\alpha/2}(t,u), \quad \widehat{\rho}^*_{1-\alpha/2}(t,u) \right],
\end{align*}
where $\widehat{\beta}^*_{\alpha/2}(t,s)$ and $\widehat{\beta}^*_{1-\alpha/2}(t,s)$ are the $(\alpha/2)$\textsuperscript{th} and $(1-\alpha/2)$\textsuperscript{th} quantiles of the bootstrap samples for $\beta(t,s)$, and similarly for $\rho(t,u)$.

To evaluate the reliability of the confidence intervals, we calculate the coverage probability deviance (CPD), which is the absolute difference between the nominal and empirical coverage probabilities, and the interval score values (score), which evaluate the coverage probability and width of the CIs simultaneously, as follows:
\begin{align*}
\text{CPD}(\beta) = \ & \Bigg\vert (1-\alpha) - \frac{1}{M G} \sum_{\iota = 1}^M \sum_{r = 1}^G \mathbb{1} \left\lbrace\widehat{\beta}^*_{\alpha/2}(t_\iota,s_r) \leq  \beta(t_\iota,s_r) \leq \widehat{\beta}^*_{1-\alpha/2}(t_\iota,s_r) \right\rbrace \Bigg\vert, \\
\text{score}(\beta) = \ & \frac{1}{M G} \sum_{\iota = 1}^M \sum_{r = 1}^G \Bigg \vert \{\widehat{\beta}^*_{1-\alpha/2}(t_\iota,s_r) - \widehat{\beta}^*_{\alpha/2}(t_\iota,s_r) \} \\
&+ \frac{2}{\alpha} \left\lbrace \widehat{\beta}^*_{\alpha/2}(t_\iota,s_r) - \beta(t_\iota,s_r) \right\rbrace \mathbb{1} \left\lbrace \beta(t_\iota,s_r) < \widehat{\beta}^*_{\alpha/2}(t_\iota,s_r) \right\rbrace \\
&+ \frac{2}{\alpha} \left\lbrace \beta(t_\iota,s_r) - \widehat{\beta}^*_{1-\alpha/2}(t_\iota,s_r) \right\rbrace \mathbb{1} \left\lbrace \beta(t_\iota,s_r) > \widehat{\beta}^*_{1-\alpha/2}(t_\iota,s_r) \right\rbrace \Bigg \vert
\end{align*}
where $\mathbb{1}\{\cdot\}$ is the indicator function. Similarly, the CPD and score metrics can be defined for the spatial autocorrelation parameter $\rho(t,u)$ following the same structure as for $\beta(t,s)$. 

In the Monte Carlo experiments, we conducted a total of 250 simulations for each combination of training sample size and spatial autocorrelation level to evaluate the performance of the proposed SFoFR model. For each simulation, $B = 199$ bootstrap samples are generated to construct 95\% (that is, $\alpha = 0.05$) confidence intervals for $\beta(t,s)$ and $\rho(t,u)$.

Our findings from the Monte Carlo experiments are provided in Table~\ref{tab:tab_1}. Our findings demonstrate that PSFoFR achieves the lowest estimation errors, the highest predictive accuracy, and the most reliable confidence intervals across all levels of spatial dependence and sample sizes. The advantages of penalization (compared to SFoFR) and accounting for spatial autocorrelation (compared to pffr) are particularly evident in our results.
\begin{small}
\begin{center}
\tabcolsep 0.1in
\renewcommand{\arraystretch}{0.92}
\begin{longtable}{@{}cccccccccc@{}} 
\caption{Computed mean $\text{RRISPEE}(\widehat{\beta})$, $\text{RRISPEE}(\widehat{\rho})$, RMSPE, $\text{CPD}(\beta)$, $\text{CPD}(\rho)$, $\text{score}(\beta)$, and $\text{score}(\rho)$ values with their standard errors (given in brackets) over 250 Monte-Carlo replications. The results are obtained under three sample sizes ($n_{\text{train}}$).}\label{tab:tab_1} \\
\toprule
{$\eta$} & {$n_{\text{train}}$} & Method & $\text{RRISPEE}(\widehat{\beta})$ & $\text{RRISPEE}(\widehat{\rho})$ & RMSPE & $\text{CPD}(\beta)$ & $\text{CPD}(\rho)$ & $\text{score}(\beta)$ & $\text{score}(\rho)$ \\ \midrule
\endfirsthead
\toprule
{$\eta$} & {$n_{\text{train}}$} & Method & $\text{RRISPEE}(\widehat{\beta})$ & $\text{RRISPEE}(\widehat{\rho})$ & RMSPE & $\text{CPD}(\beta)$ & $\text{CPD}(\rho)$ & $\text{score}(\beta)$ & $\text{score}(\rho)$ \\ \midrule
\endhead
\midrule
\multicolumn{10}{r}{Continued on next page} \\ 
\endfoot
\endlastfoot
0.1 & 100   & pffr  & $2.700$ & $--$ & $4.269$ & $0.346$ & $--$ & $0.918$ & $--$ \\
    &       &       & ($0.592$) & ($--$) & ($1.743$) & ($0.101$) & ($--$) & ($0.384$) & ($--$)  \\
    &       & SFoFR   & $11.441$ & $30.312$ & $18.359$ & $0.425$ & $0.223$ & $5.294$ & $0.236$ \\
    &       &       & ($4.633$) & ($19.617$) & ($2.672$) & ($0.182$) & ($0.180$) & ($4.070$) & ($0.113$) \\
    &       & PSFoFR & $0.132$ & $10.797$ & $0.088$ & $0.015$ & $0.152$ & $0.022$ & $0.083$ \\
    &       &       & ($0.015$) & ($2.369$) & ($0.007$) & ($0.010$) & ($0.115$) & ($0.002$) & ($0.021$) \\
\cmidrule(l){2-10}				
    & 250   & pffr  & $2.665$ & $--$ & $3.190$ & $0.574$ & $--$ & $1.349$ & $--$ \\
    &       &       & ($0.309$) & ($--$) & ($0.581$) & ($0.030$) & ($--$) & ($0.227$) & ($--$)  \\
    &       & SFoFR   & $8.905$ & $21.997$ & $16.402$ & $0.341$ & $0.289$ & $3.079$ & $0.267$ \\
    &       &       & ($1.456$) & ($1.961$) & ($2.105$) & ($0.140$) & ($0.167$) & ($1.497$) & ($0.090$) \\
    &       & PSFoFR & $0.090$ & $8.818$ & $0.056$ & $0.009$ & $0.157$ & $0.015$ & $0.072$ \\
    &       &       & ($0.008$) & ($0.612$) & ($0.003$) & ($0.009$) & ($0.161$) & ($0.001$) & ($0.032$) \\
\cmidrule(l){2-10}
& 500   & pffr  & $2.781$ & $--$ & $2.938$ & $0.641$ & $--$ & $1.782$ & $--$ \\
    &       &       & ($0.190$) & ($--$) & ($0.308$) & ($0.020$) & ($--$) & ($0.133$) & ($--$)  \\
    &       & SFoFR   & $8.043$ & $21.390$ & $16.428$ & $0.272$ & $0.441$ & $2.376$ & $0.288$ \\
    &       &       & ($1.222$) & ($1.656$) & ($1.100$) & ($0.137$) & ($0.139$) & ($0.848$) & ($0.088$) \\
    &       & PSFoFR & $0.072$ & $8.770$ & $0.040$ & $0.010$ & $0.231$ & $0.012$ & $0.087$ \\
    &       &       & ($0.004$) & ($1.290$) & ($0.002$) & ($0.009$) & ($0.198$) & ($0.001$) & ($0.038$) \\
\midrule

0.5 & 100   & pffr  & $5.687$ & $--$ & $21.941$ & $0.029$ & $--$ & $0.812$ & $--$ \\
    &       &       & ($1.573$) & ($--$) & ($5.884$) & ($0.019$) & ($--$) & ($0.222$) & ($--$)  \\
    &       & SFoFR   & $10.051$ & $20.552$ & $17.930$ & $0.409$ & $0.522$ & $3.885$ & $1.625$ \\
    &       &       & ($2.256$) & ($1.133$) & ($1.661$) & ($0.132$) & ($0.108$) & ($1.681$) & ($0.482$) \\
    &       & PSFoFR & $0.133$ & $8.069$ & $0.103$ & $0.036$ & $0.281$ & $0.038$ & $0.454$ \\
    &       &       & ($0.012$) & ($0.590$) & ($0.012$) & ($0.005$) & ($0.206$) & ($0.009$) & ($0.244$) \\
\cmidrule(l){2-10}
					
& 250   & pffr  & $3.666$ & $--$ & $17.903$ & $0.043$ & $--$ & $0.512$ & $--$ \\
    &       &       & ($1.081$) & ($--$) & ($3.028$) & ($0.030$) & ($--$) & ($0.142$) & ($--$)  \\
    &       & SFoFR   & $8.629$ & $19.802$ & $16.524$ & $0.443$ & $0.607$ & $3.747$ & $1.890$ \\
    &       &       & ($1.291$) & ($0.593$) & ($1.309$) & ($0.175$) & ($0.165$) & ($1.187$) & ($0.544$) \\
    &       & PSFoFR & $0.090$ & $7.538$ & $0.074$ & $0.031$ & $0.219$ & $0.025$ & $0.346$ \\
    &       &       & ($0.007$) & ($0.734$) & ($0.010$) & ($0.010$) & ($0.183$) & ($0.005$) & ($0.213$) \\
\cmidrule(l){2-10}

& 500   & pffr  & $3.622$ & $--$ & $16.279$ & $0.159$ & $--$ & $0.638$ & $--$ \\
    &       &       & ($1.082$) & ($--$) & ($2.046$) & ($0.100$) & ($--$) & ($0.263$) & ($--$)  \\
    &       & SFoFR   & $8.506$ & $19.723$ & $16.520$ & $0.503$ & $0.666$ & $4.089$ & $2.086$ \\
    &       &       & ($1.397$) & ($0.583$) & ($1.108$) & ($0.127$) & ($0.110$) & ($1.244$) & ($0.451$) \\
    &       & PSFoFR & $0.071$ & $6.830$ & $0.057$ & $0.026$ & $0.221$ & $0.016$ & $0.284$ \\
    &       &       & ($0.006$) & ($0.689$) & ($0.008$) & ($0.009$) & ($0.125$) & ($0.003$) & ($0.070$) \\
\midrule

0.9 & 100 & pffr & $12.440$ & $--$ & $148.602$ & $0.047$ & $--$ & $1.923$ & $--$ \\
    &       &       & ($6.638$) & ($--$) & ($88.508$) & ($0.018$) & ($--$) & ($0.616$) & ($--$)  \\
    &       & SFoFR   & $11.405$ & $19.424$ & $33.625$ & $0.528$ & $0.612$ & $5.910$ & $3.352$ \\
    &       &       & ($2.350$) & ($0.275$) & ($11.418$) & ($0.127$) & ($0.124$) & ($1.957$) & ($0.864$) \\
    &       & PSFoFR & $0.203$ & $7.810$ & $0.629$ & $0.032$ & $0.117$ & $0.073$ & $0.561$ \\
    &       &       & ($0.070$) & ($0.589$) & ($0.254$) & ($0.016$) & ($0.104$) & ($0.028$) & ($0.192$) \\
\cmidrule(l){2-10}
					
& 250   & pffr & $8.381$ & $--$ & $125.278$ & $0.046$ & $--$ & $1.161$ & $--$ \\
    &       &       & ($3.599$) & ($--$) & ($76.351$) & ($0.025$) & ($--$) & ($0.412$) & ($--$)  \\
    &       & SFoFR   & $10.125$ & $19.328$ & $21.722$ & $0.577$ & $0.679$ & $5.635$ & $3.810$ \\
    &       &       & ($1.382$) & ($0.242$) & ($3.981$) & ($0.135$) & ($0.127$) & ($1.609$) & ($0.890$) \\
    &       & PSFoFR & $0.200$ & $7.064$ & $0.594$ & $0.068$ & $0.137$ & $0.064$ & $0.452$ \\
    &       &       & ($0.151$) & ($0.443$) & ($0.172$) & ($0.090$) & ($0.084$) & ($0.058$) & ($0.080$) \\
\cmidrule(l){2-10}

& 500   & pffr  & $5.909$ & $--$ & $93.557$ & $0.073$ & $--$ & $0.925$ & $--$ \\
    &       &       & ($2.069$) & ($--$) & ($52.773$) & ($0.060$) & ($--$) & ($0.320$) & ($--$)  \\
    &       & SFoFR   & $9.690$ & $19.395$ & $18.337$ & $0.692$ & $0.783$ & $6.701$ & $4.489$ \\
    &       &       & ($0.973$) & ($0.250$) & ($2.270$) & ($0.086$) & ($0.095$) & ($1.008$) & ($0.735$) \\
    &       & PSFoFR & $0.119$ & $6.418$ & $0.594$ & $0.065$ & $0.124$ & $0.043$ & $0.420$ \\
    &       &       & ($0.097$) & ($0.440$) & ($0.176$) & ($0.079$) & ($0.075$) & ($0.033$) & ($0.045$) \\
\bottomrule
\end{longtable}
\end{center}
\end{small}

\vspace{-.2in}

When examining the estimation accuracy of the bivariate regression function $\beta(t,s)$, our results show that the proposed PSFoFR method consistently outperforms both SFoFR and pffr, as reflected in its substantially lower RRISPEE values. The penalization in PSFoFR plays a crucial role in controling the complexity of the model and mitigating overfitting, which is a key limitation of SFoFR. Although SFoFR accounts for spatial autocorrelation, its lack of penalization leads to inflated estimation errors. For example, under weak spatial dependence ($\eta = 0.1$), SFoFR produces an RRISPEE($\widehat{\beta}$) of 11.441, whereas PSFoFR dramatically reduces this error to 0.132, highlighting the effectiveness of penalization. Even as sample sizes increase, SFoFR continues to exhibit higher estimation errors compared to PSFoFR, reinforcing the need for regularization in functional regression with spatial dependence. In contrast, while pffr benefits from penalization, it fails to model spatial dependence, leading to moderate estimation errors. This issue becomes more pronounced in high spatial correlation settings ($\eta = 0.9)$, where pffr’s RRISPEE($\widehat{\beta}$) rises to 12.440, while PSFoFR maintains a much higher accuracy with an RRISPEE($\widehat{\beta}$) of 0.203. These findings confirm that both penalization and spatial autocorrelation modeling are necessary for optimal estimation performance, and PSFoFR integrates both aspects.

When evaluating the estimation accuracy of the spatial autocorrelation function $\rho(t,u)$, which is only estimated by SFoFR and PSFoFR, the results show that PSFoFR provides much more accurate estimates of $\rho(t,u)$ than SFoFR, with significantly lower RRISPEE($\widehat{\rho}$) values across all settings. For example, in the case of moderate spatial dependence, ($\eta = 0.5)$, SFoFR produces an RRISPEE ($\widehat{\rho}$) of 19.802, while PSFoFR achieves a dramatically lower 7.538, demonstrating that penalization improves the stability and accuracy of spatial correlation estimation. This trend persists across different levels of spatial dependence, confirming that PSFoFR effectively balances spatial autocorrelation modeling with regularization, ensuring more precise parameter estimation.

The predictive accuracy, measured by RMSPE, underscores the superiority of PSFoFR. PSFoFR consistently achieves the lowest RMSPE across all sample sizes and spatial dependence levels, demonstrating its ability to generalize well to new data. In contrast, SFoFR, despite modeling spatial correlation, exhibits much higher RMSPE values due to overfitting caused by the lack of penalization. For example, at $\eta = 0.1$ and $n_{\text{train}} = 100$, SFoFR produces an RMSPE of 18.359, while PSFoFR reduces this to a remarkably low 0.088. Even more striking is the failure of pffr in highly spatially dependent settings ($\eta = 0.9)$, where its RMSPE soars to 148.602, compared to just 0.629 for PSFoFR. This significant discrepancy highlights that ignoring spatial dependencies severely degrades predictive performance, reinforcing the need to incorporate spatial autocorrelation in functional regression.

From Table~\ref{tab:tab_1},  PSFoFR provides the most reliable confidence intervals, with lower CPD values (indicating better nominal coverage) and lower interval score values (indicating narrower, more informative intervals). In contrast, SFoFR exhibits higher CPD and score values, reflecting wider and less reliable intervals due to the lack of penalization. For example, at $\eta = 0.5$, SFoFR’s CPD($\rho$) is 0.607, while PSFoFR reduces it to 0.219, confirming that penalization stabilizes uncertainty quantification in spatial functional models. An important nuance in our inferential results concerns the CPD for the spatial autocorrelation function, $\rho(t,u)$. The observed discrepancy between nominal and empirical coverage probabilities for the confidence intervals of $\rho(t,u)$, as measured by CPD, may stem from finite-sample bias introduced by the penalization step. While our asymptotic results establish $\sqrt{n}$-consistency and normality, finite-sample performance can be affected by this bias. This bias affects the bootstrap-based confidence intervals, leading to under-coverage (empirical coverage ~0.75–0.85 for nominal 0.95 in some cases). De-biasing procedures, analogous to those proposed by \cite{Geer2014} for high-dimensional models, could be adapted to this functional setting to improve inference, particularly for the spatial lag parameter $\rho(t,u)$. However, bias correction for spatial lag parameters in functional models remains an underdeveloped area and is a promising direction for future research.

A notable and important result from Table~\ref{tab:tab_1} is that while SFoFR accounts for spatial dependence and shows improved predictive performance over pffr under strong spatial dependence (as seen in lower RMSPE for $\eta = 0.9$), its estimation accuracy is hindered by the challenges in selecting the optimal number of principal components, which can lead to under-fitting or over-fitting. In other words, the SFoFR method is consistently outperformed by pffr in estimating the regression coefficient function $\beta(t,s)$, as measured by RRISPEE($\widehat{\beta}$). This is consistent with the bias-variance trade-off, where improper component selection increases variance or bias \citep{Crainiceanu2009, ivanescu2015}. The pffr model is misspecified for these spatially correlated data, which introduces bias into its estimate of $\beta(t,s)$. However, its use of a strong roughness penalty provides powerful regularization, resulting in an estimator with very low variance. In contrast, the SFoFR model is correctly specified and thus has a low bias. Because SFoFR is an unpenalized method, it is prone to overfitting the sample data, leading to an estimator with high variance. In these simulations, the variance inflation from SFoFR's lack of regularization is more detrimental to its overall error than the bias from pffr's model misspecification. This finding strongly motivates the need for our proposed PSFoFR model. By integrating both a correct spatial autoregressive structure (to ensure low bias) and an explicit roughness penalty (to ensure low variance), PSFoFR successfully navigates this trade-off, leading to its superior estimation accuracy over both alternatives.  

\begin{figure}[!htb]
\centering
\includegraphics[width=8.2cm]{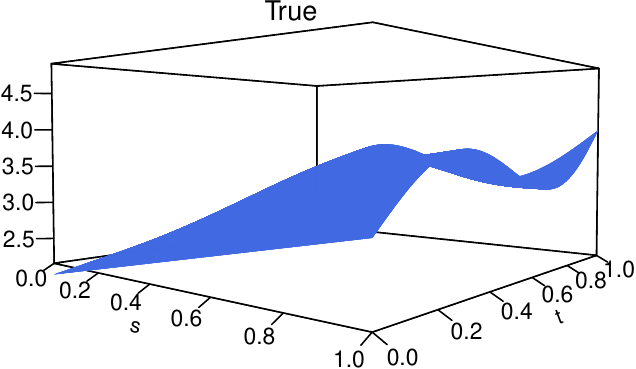}
\qquad
\includegraphics[width=8.2cm]{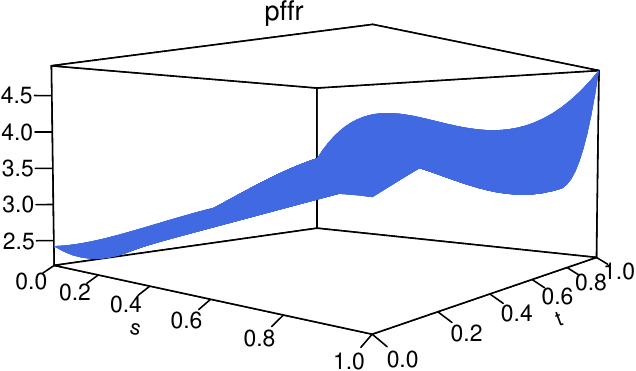}
\\
\includegraphics[width=8.2cm]{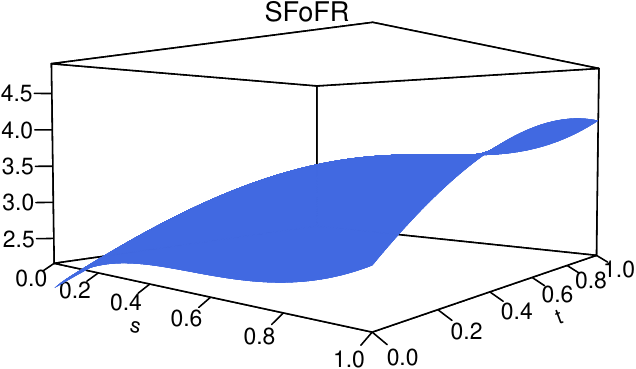}
\qquad
\includegraphics[width=8.2cm]{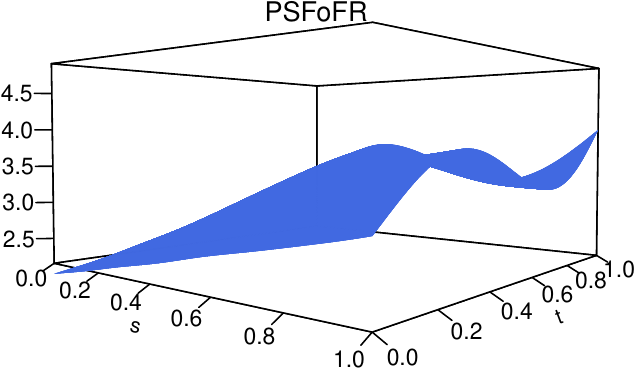}
\caption{Three-dimensional representation of the true regression coefficient function $\beta(t,s)$ (True) and the estimated regression coefficient functions by the methods (pffr, SFoFR, and PSFoFR). The plots are generated for a sample size of $n_{\text{train}} = 500$ with $\eta = 0.5$.}\label{fig:Fig_1}
\end{figure}

Figure~\ref{fig:Fig_1} illustrates the estimated regression coefficient function $\beta(t,s)$ (from a single experiment) obtained from each method compared to the true coefficient function. The PSFoFR method produces an estimate that closely aligns with the true function, maintains smoothness, and captures the essential structural features of the data. In contrast, SFoFR, despite modeling spatial dependence, exhibits noticeable deviations from the true function, likely due to the lack of penalization, which results in overfitting and increased variance. The pffr method, while benefiting from penalization, fails to account for spatial dependence, leading to systematic distortions in its estimates, particularly in regions of high spatial correlation.

\begin{figure}[!htb]
\centering
\includegraphics[width=5.75cm]{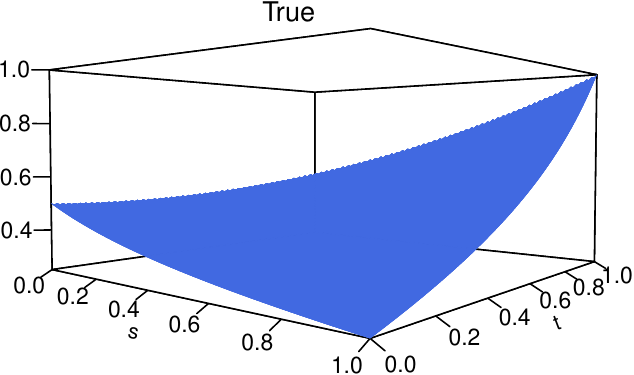}
\quad
\includegraphics[width=5.75cm]{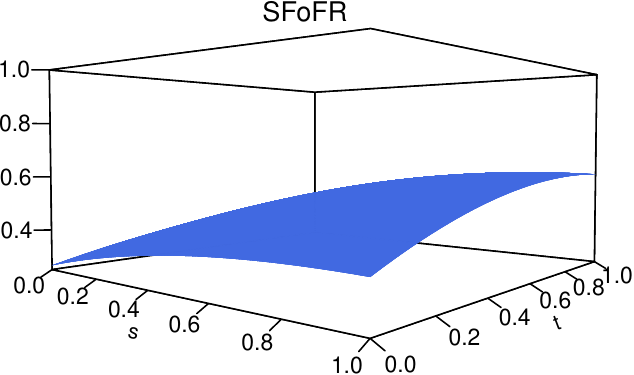}
\quad
\includegraphics[width=5.75cm]{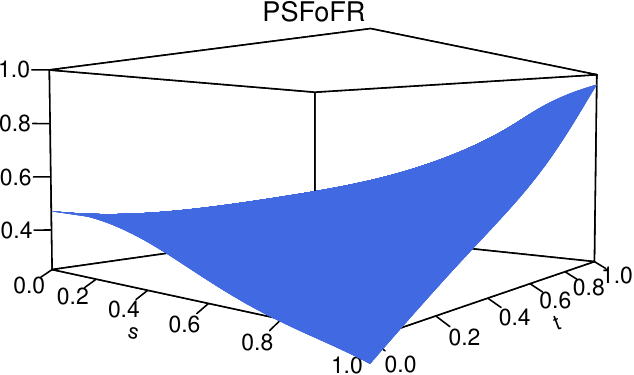}
\caption{Three-dimensional representation of the true spatial autocorrelation function $\rho(t,u)$ (left panel) and the estimated spatial autocorrelation functions by SFoFR (middle panel) and PSFoFR (right panel). The plots are generated for a sample size of $n_{\text{train}} = 500$ with $\eta = 0.5$.}\label{fig:Fig_2}
\end{figure}

Figure~\ref{fig:Fig_2} displays the estimated spatial autocorrelation function $\rho(t,u)$ (from a single experiment), comparing the results of SFoFR and PSFoFR with the true function. The PSFoFR method significantly outperforms SFoFR, as it produces estimates with lower error and a more stable spatial structure. Penalization in PSFoFR reduces the variance of the estimation while still preserving the key spatial characteristics of $\rho(t,u)$. In contrast, SFoFR’s estimate shows excessive instability, reinforcing the need for penalization to improve robustness in spatial functional regression.

\begin{figure}[!htb]
\centering
\includegraphics[width=9cm]{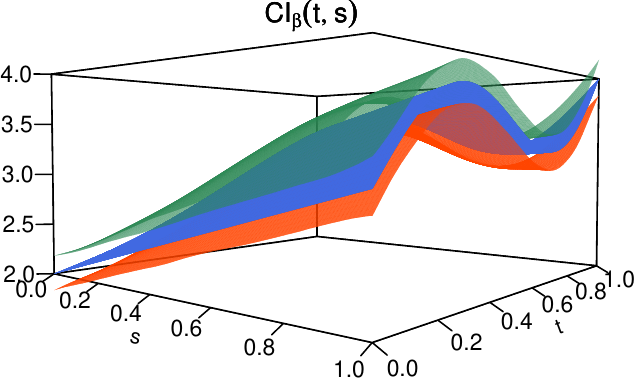}
\quad
\includegraphics[width=9cm]{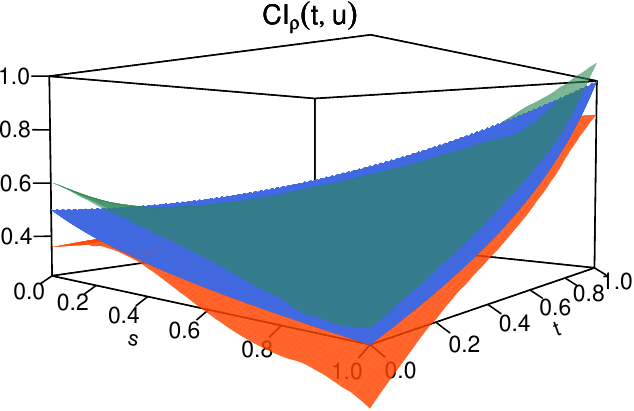}
\caption{Three-dimensional representation of the true regression coefficient function $\beta(t,s)$ (left panel) and the spatial autocorrelation function $\rho(t,u)$  (right panel), along with their confidence interval bounds constructed using the proposed method. The true functions are depicted in royal blue, while the lower and upper confidence bounds are shown in orange and sea green, respectively. Transparency is applied to the confidence bounds to improve visibility and distinguishability. The plots are generated for a sample size of $n_{\text{train}} = 500$ with $\eta = 0.5$.}\label{fig:Fig_3}
\end{figure}

Figure~\ref{fig:Fig_3} presents the confidence interval bounds for $\beta(t,s)$ and $\rho(t,u)$ (from a single experiment) under PSFoFR. Note that both the pffr and SFoFR methods are unable to provide accurate confidence intervals for $\beta(t,s)$ and $\rho(t,u)$, and thus their results are not provided. The proposed method achieves narrow, well-calibrated confidence intervals that closely follow the true functions, highlighting its reliability for uncertainty quantification. The tight bounds indicate that PSFoFR produces stable and efficient parameter estimates, reducing overfitting while maintaining flexibility in capturing functional dependencies.

While our proposed PSFoFR method demonstrates superior statistical performance, it is also important to consider its computational cost. To this end, we recorded the elapsed wall-clock time required to fit a single Monte Carlo replication for each method. The simulations were conducted on a Dell PC with a 13th Gen Intel(R) Core(TM) i9-13900HX processor and 64 GB of RAM. The results are summarized in Table~\ref{tab:tab_2}. The results show that PSFoFR is the most computationally intensive method, followed by pffr, with SFoFR being the fastest. This hierarchy is expected and directly reflects the complexity of the estimation procedures. The SFoFR method's speed stems from its direct, non-iterative estimation on a reduced set of principal component scores. In contrast, both pffr and PSFoFR require an automated grid search to find the optimal smoothing parameters by minimizing an information criterion, which is the primary driver of their computational cost. Our proposed PSFoFR is the most demanding because, in addition to this optimization step, it incorporates the matrix operations required for the two-stage least-squares procedure (i.e., constructing instrumental variables and performing projections). 

\begin{table}[!htb]
\centering
\tabcolsep 0.15in
\caption{\small{Elapsed computing times for the pfr, SFoFR, and PSFoFR (in seconds). The computing times are obtained from a single Monte-Carlo experiment when $n_{\text{train}} \in \{100, 250, 500\}$ and $\eta \in \{0.1, 0.5, 0.9\}$.}}
\label{tab:tab_2}
\begin{tabular}{@{}lccccccccc@{}}
\toprule
& \multicolumn{3}{c}{$\eta = 0.1$} & \multicolumn{3}{c}{$\eta = 0.5$} & \multicolumn{3}{c}{$\eta = 0.9$} \\
\midrule
$n_{\text{train}}$ & pfr & SFoFR & PSFoFR & pfr & SFoFR & PSFoFR & pfr & SFoFR & PSFoFR\\
\midrule
100 & 2.64 & 0.19 & 6.04 & 2.57 & 0.15 & 6.17 & 2.58 & 0.17 & 6.16 \\
250 & 5.36 & 0.21 & 14.09 & 5.41 & 0.20 & 15.01 & 5.51 & 0.23 & 15.06 \\
500 & 9.74 & 3.51 & 27.84 & 9.93 & 3.55 & 27.88 & 9.91 & 3.62 & 28.03 \\
\bottomrule
\end{tabular}
\end{table}

\section{North Dakota weather data analysis}\label{sec:5}

\begin{figure}[!htb]
\centering
\includegraphics[height=7.9cm]{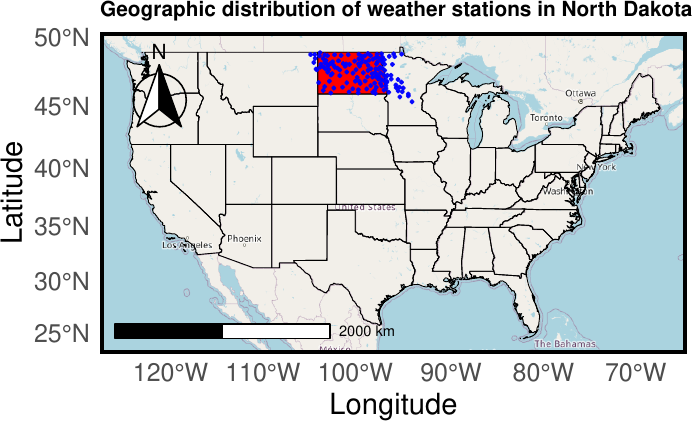}
\caption{Spatial distribution of weather stations in North Dakota.}
\label{fig:Fig_4}
\end{figure}

To illustrate the practical utility of our proposed PSFoFR model, we apply it to daily weather data collected from the North Dakota Agricultural Weather Network (NDAWN) for the entire year 2024 (the dataset is publicly available at \url{https://ndawn.ndsu.nodak.edu/}). As shown in Figure~\ref{fig:Fig_4}, the data set consists of observations from 140 weather stations spread throughout North Dakota, with station selection based on data availability.

Each station provides functional measurements of key meteorological variables over time, enabling a comprehensive investigation of temporal and spatial dependencies in weather patterns. In this study, we focus on analyzing the relationship between wind chill as a functional response and wind speed as a functional predictor, both measured on a daily basis. 

\begin{figure}[!htb]
\centering
\includegraphics[width=9cm]{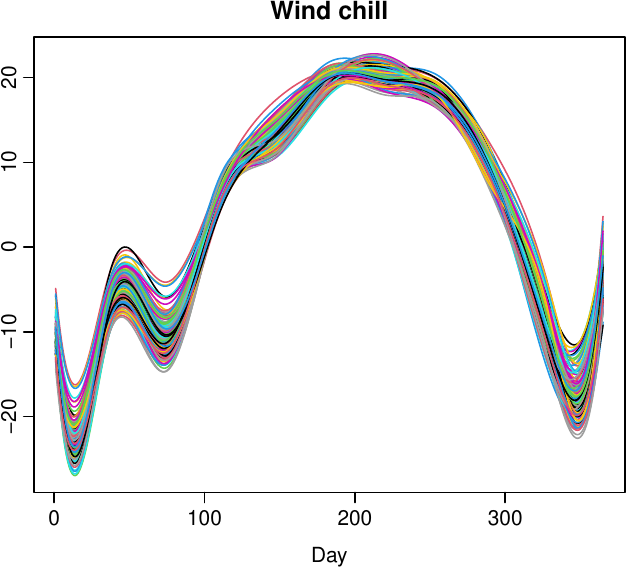}
\quad
\includegraphics[width=9cm]{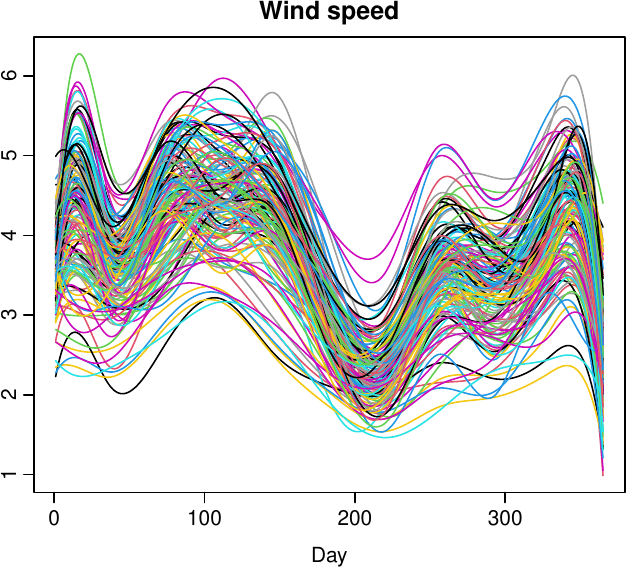}
\caption{Graphical representation of wind chill (left panel) and wind speed (right panel) for 2024, based on data collected from 140 weather stations in North Dakota. Different colors indicate different weather stations, with observations recorded as functions of days $(1 \leq t,s \leq 365)$.}\label{fig:Fig_5}
\end{figure}

Wind chill, a key metric in meteorology and public safety, quantifies perceived temperature by factoring in wind speed and ambient air temperature. In cold regions like North Dakota, extreme wind chills increase the risk of hypothermia and frostbite while affecting agriculture, energy, and emergency planning. Higher wind speeds accelerate heat loss, making temperatures feel much colder. Traditional wind chill indices assume a static relationship, but the influence of wind speed varies across locations and time. By modeling wind speed as a functional predictor, our approach captures its dynamic effects, offering a more precise representation of wind chill in diverse regions. In Figure~\ref{fig:Fig_5}, we display a graphical representation of the wind chill and wind speed variables collected from 140 weather stations in North Dakota; the raw data were smoothed using 13 cubic B-spline basis functions for improved visualization.

A major challenge in modeling meteorological data is spatial dependence-weather stations are highly correlated because of shared climatic influences and regional wind patterns. Ignoring this can bias estimates and weaken predictions. The PSFoFR model tackles this by integrating a spatial autoregressive component, capturing interactions between wind chill observations across locations. This enhances accuracy, ensuring more reliable and generalizable forecasts.

To construct the spatial weight matrix $\bm{W} = (w_{ij})_{140 \times 140}$ for the 140 weather stations in North Dakota, we utilize a $K$-nearest neighbors approach with a bi-square weight function and adaptive bandwidth. The geographic coordinates (longitude and latitude) of the stations are first used to compute the pairwise great-circle distances $\{d_{ij}\}$ between the locations $i$ and $j$. For each station $i$, we determine its $h = 4$ nearest neighbors based on these distances ($h$ is determined by cross-validation). The bandwidth $H_i$ is adaptively selected as the maximum distance among the four nearest neighbors of the station $i$, that is, $H_i = \max \{ d_{ij} : j \in N_4(i) \}$, where $N_4(i)$ represents the set of the four closest stations to $i$. The spatial weight for each neighbor $j$ of station $i$ is then computed using the bi-square kernel function:
\begin{equation*}
w_{ij} = \left( 1 - \left( \frac{d_{ij}}{H_i} \right)^2 \right)^2, \quad \text{for } j \in N_4(i).
\end{equation*}
Locations that are not among the four nearest neighbors receive a weight of zero ($w_{ij} = 0$ for $j \notin N_4(i)$). To ensure that the weights sum to one for each station, they are normalized as $w_{ij} = \frac{w_{ij}}{\sum_{j \in N_4(i)} w_{ij}}.$

We compute the functional Moran’s I statistic \citep{Khoo2023} to assess the degree of spatial autocorrelation in the functional response variable. To achieve this, we represent the functional response using a basis-expansion approach. Specifically, we employ a B-spline basis expansion, where the set of basis functions is denoted as $\bm{\varphi}(t) = \{\varphi_1(t), \varphi_2(t), \dots\}$, and the corresponding expansion coefficients are represented by the matrix $\bm{V} = [V_1, V_2, \dots]$. This expansion allows us to express the functional response in a low-dimensional space while preserving its essential characteristics. Using this representation, the functional Moran’s I statistic is defined as:
\begin{equation*}
I[\Y(t)] = \frac{\bm{\varphi}^\top(t) \bm{V}^\top \bm{W} \bm{V} \bm{\varphi}(t)}{\bm{\varphi}^\top(t) \bm{V}^\top \bm{V} \bm{\varphi}(t)},
\end{equation*}
where $\bm{W}$ is the spatial weight matrix that captures the spatial dependencies between observations. The numerator quantifies the degree of spatial similarity in the functional response, while the denominator serves as a normalization factor. A higher value of $I[\Y(t)]$ indicates stronger spatial autocorrelation, whereas values closer to zero suggest weak or no spatial dependence. 

\begin{figure}[!htb]
\centering
\includegraphics[width=10cm]{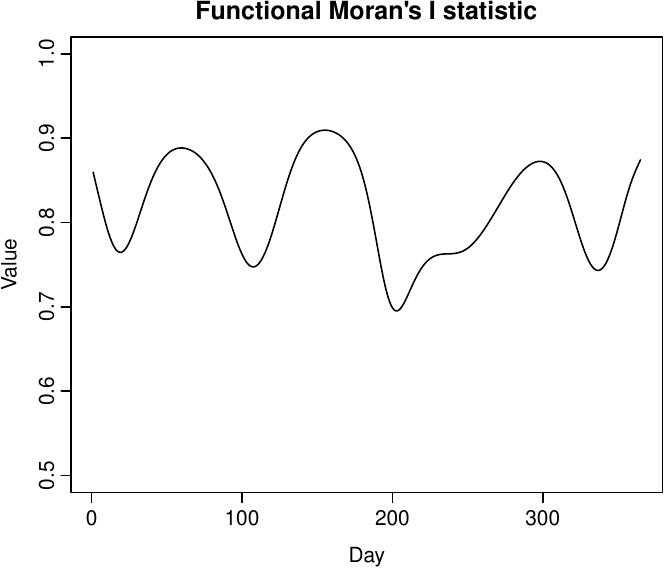}
\caption{The temporal variation of the functional Moran's I statistic, illustrating the spatial and functional characteristics of the North Dakota data.}\label{fig:Fig_4_2}
\end{figure}

The computed functional Moran’s I statistic is presented in Figure~\ref{fig:Fig_4_2}, revealing a substantial and persistent degree of spatial autocorrelation in the functional response variable across the temporal domain. The statistics values remain consistently high, fluctuating between approximately 0.7 and 0.9, indicating a strong spatial dependence throughout the year. The periodic oscillations suggest that the spatial autocorrelation structure exhibits temporal variations, potentially driven by seasonal effects or other underlying cyclic influences. These findings underscore the need to incorporate spatial dependence into the functional regression model to ensure accurate statistical inference and prediction.

Using the North Dakota weather dataset, we analyze the SFoFR model:
\begin{equation*}
\Y_i(t) = \sum_{j=1}^n w_{ij} \int_{u=1}^{365} \Y_j(u) \rho(t,u) \, du + \int_{s=1}^{365} \X_i(s) \beta(t,s) \, ds + \epsilon_i(t).
\end{equation*}
To estimate the model parameters, we employ the proposed PSFoFR, SFoFR, and pffr methods. The performance of the model is assessed using the root mean squared error (RMSE) and the coefficient of determination $(R^2)$, defined as:
\begin{align*}
\text{RMSE} &= 100 \times \sqrt{\frac{\Vert \Y(t) - \widehat{\Y}(t) \Vert_2^2}{\Vert \Y(t) \Vert_2^2}} \\
R^2 &= 1 - \frac{\int_{t=1}^{365}\left\lbrace\Y(t) - \widehat{\Y}(t)\right\rbrace^2 dt}{\int_{t=1}^{365}\left\lbrace\Y(t) - \overline{\Y}(t)\right\rbrace^2 dt},
\end{align*}
where $\widehat{\Y}(t)$ represents the estimated wind chill and $\overline{\Y}(t)$ denotes the average wind chill for the entire year 2024. 

Our results indicate that the PSFoFR model achieves the lowest RMSE (3.260), indicating the highest predictive accuracy, and the highest $R^2$ (0.999), demonstrating near-perfect goodness of fit. In comparison, the SFoFR and pffr models exhibit higher RMSE values (11.694 and 8.906, respectively) and lower $R^2$ scores (0.982 and 0.987), suggesting that they capture less variance in the observed data. These results highlight the superior effectiveness of the PSFoFR model in capturing complex spatial and functional dependencies in wind chill estimation. 

\begin{figure}[!htb]
\centering
\includegraphics[width=5.82cm]{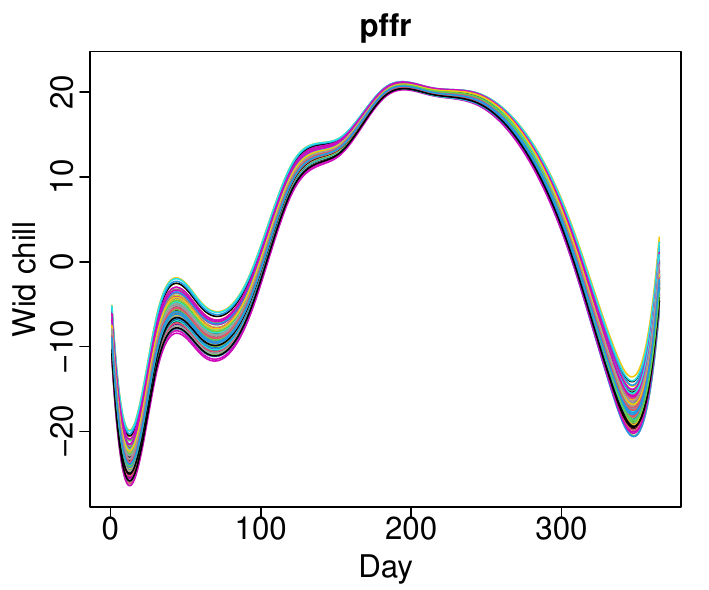}
\quad
\includegraphics[width=5.82cm]{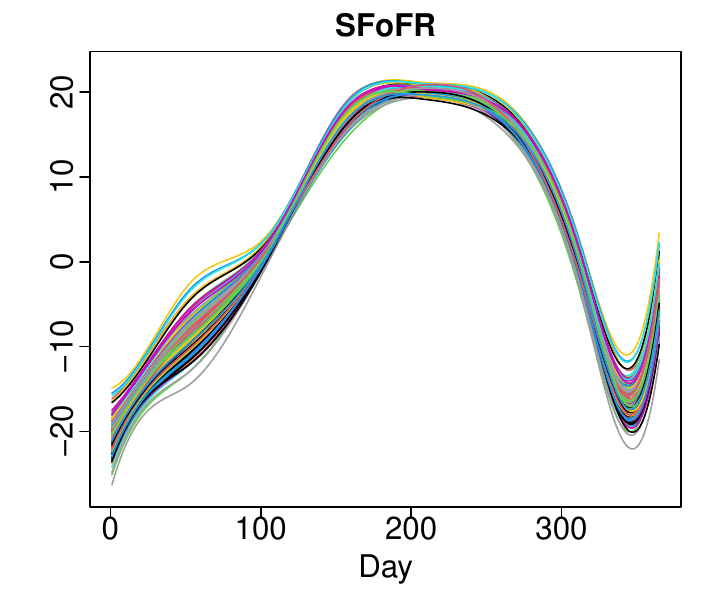}
\quad
\includegraphics[width=5.82cm]{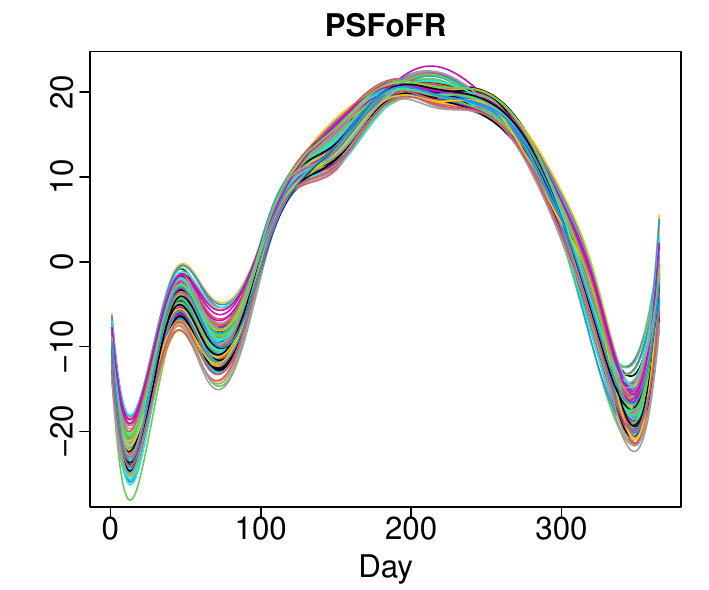}
\caption{Fitted wind chill curves for 140 weather stations in North Dakota using three different models: pffr (left panel), SFoFR (middle panel), and the proposed PSFoFR (right panel).}\label{fig:Fig_6}
\end{figure}

Figure~\ref{fig:Fig_6} provides a graphical comparison of the fitted wind chill curves obtained using different methods. From Figure~\ref{fig:Fig_5}, wind chill is most severe in mid-winter and warms toward late summer, while wind speed exhibits higher cross-station variability and intermittent peaks. This is fully consistent with the wind-chill mechanism, in which perceived temperature combines air temperature with wind speed; stronger winds enhance convective heat loss and make it feel colder. In Figure~\ref{fig:Fig_6}, all three models reproduce the dominant seasonal pattern of wind chill, but PSFoFR fits track the amplitude and phase of the curves most closely across stations, particularly around winter troughs and the late-year rebound—where spatial borrowing across nearby stations is informative. The unpenalized SFoFR is slightly more variable around peaks/troughs, while pffr occasionally underfits stations with atypical local wind regimes, a plausible outcome when spatial dependence is ignored in a region with coherent winter wind fields.

\begin{figure}[!htb]
\centering
\includegraphics[width=5.75cm]{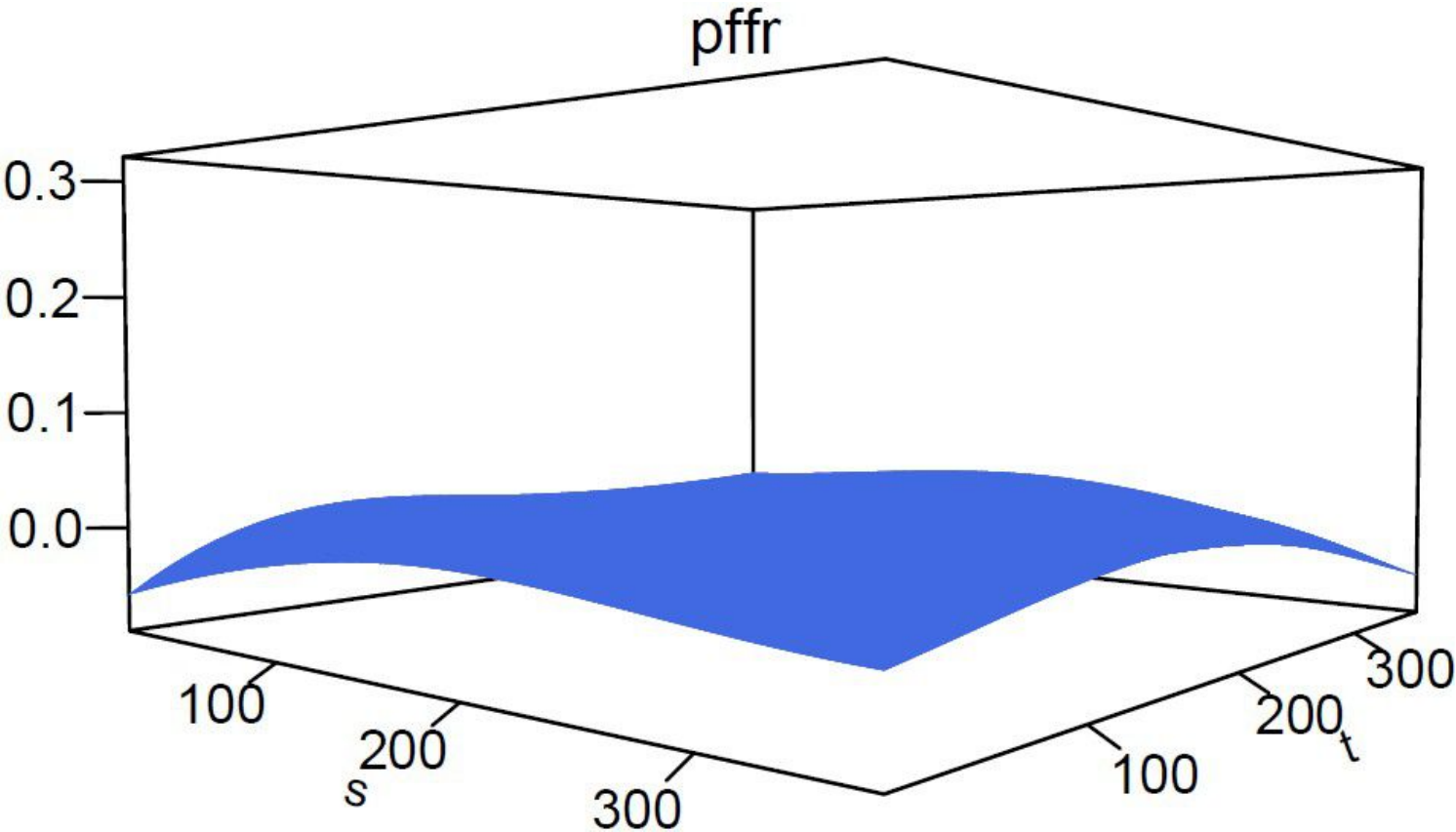}
\quad
\includegraphics[width=5.75cm]{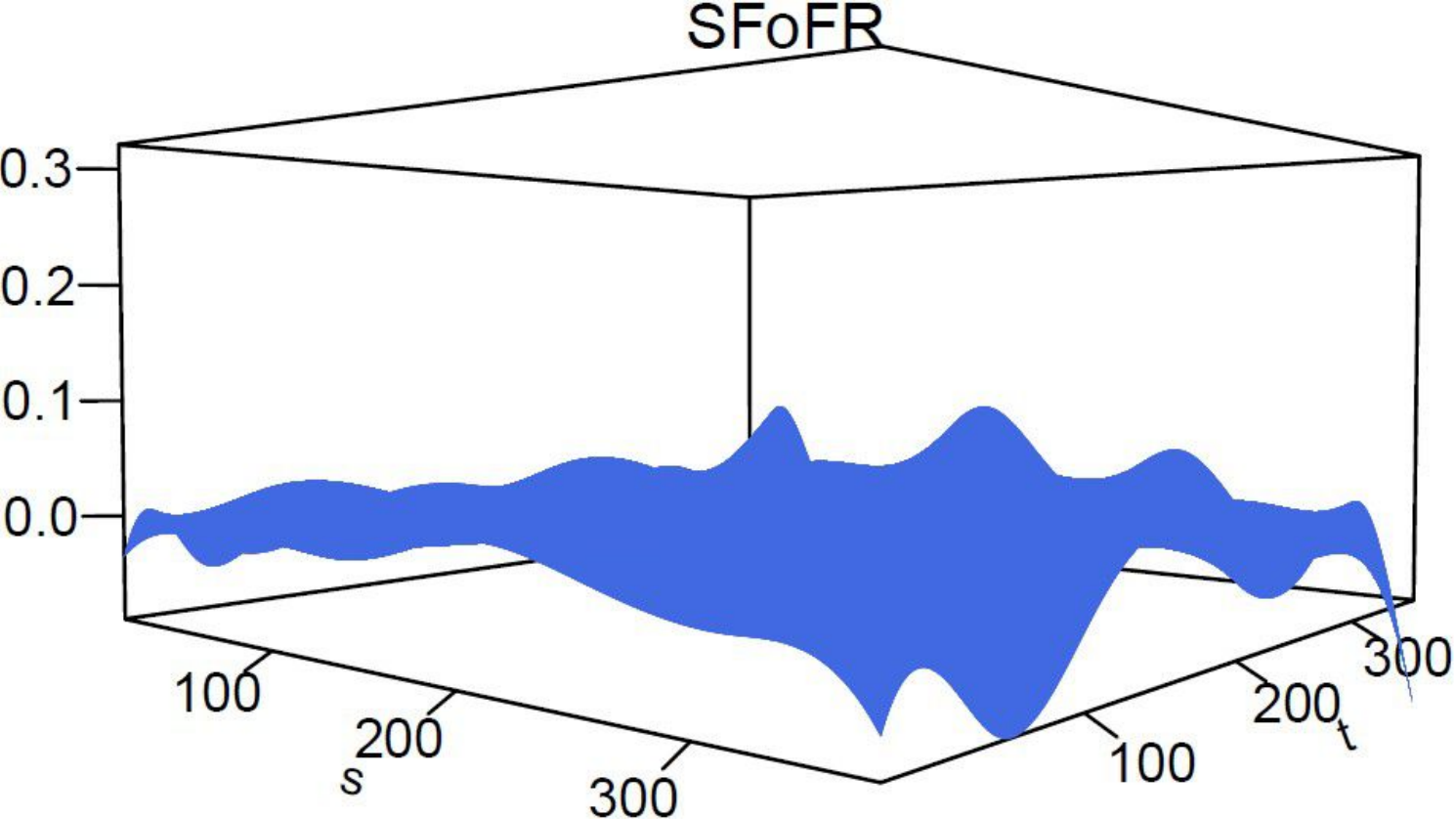}
\quad
\includegraphics[width=5.75cm]{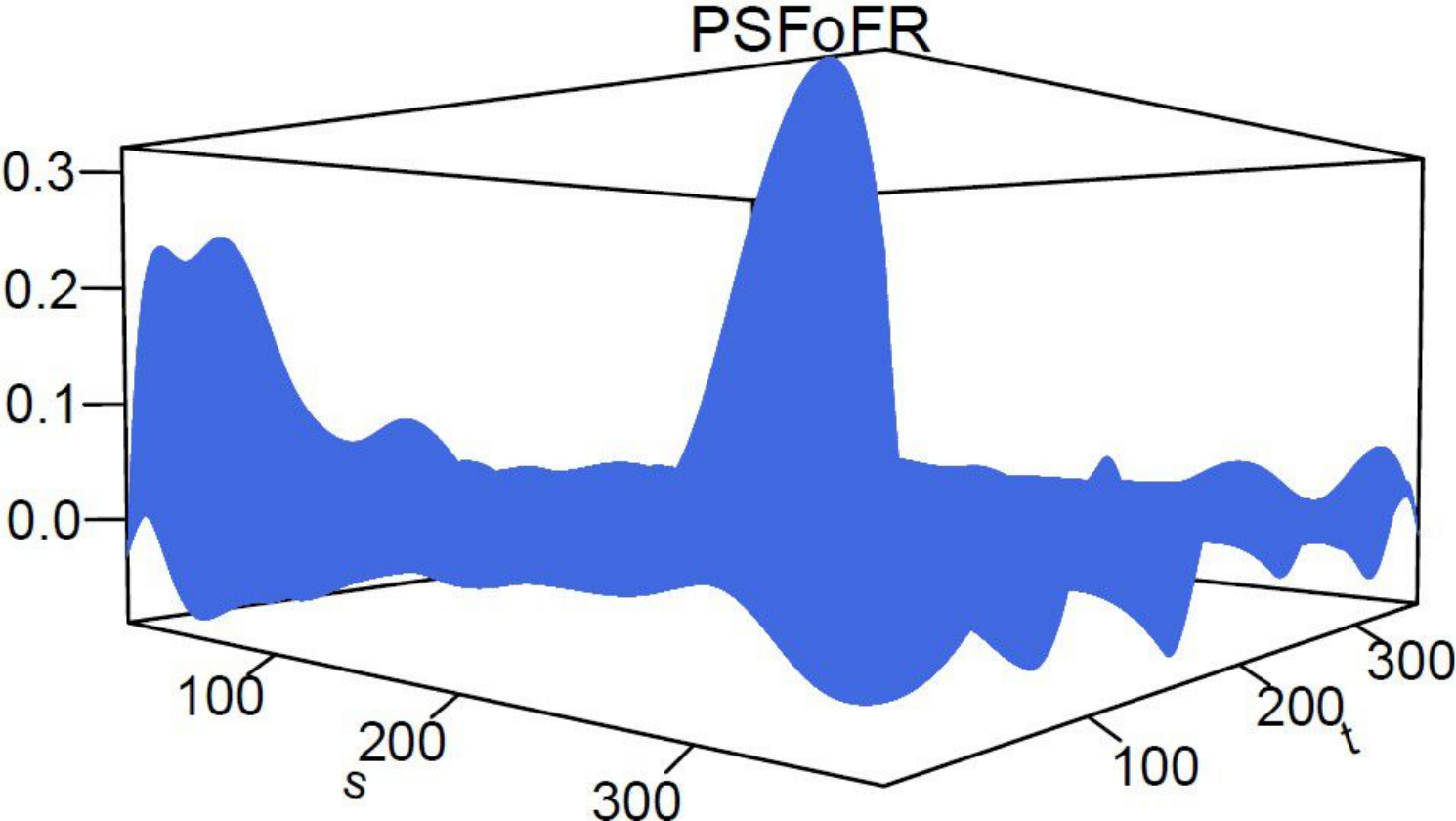}
\caption{Three-dimensional representation of the estimated regression coefficient function $\beta(t,s)$ by pffr (left panel), SFoFR (middle panel), and PSFoFR (right panel).}\label{fig:Fig_7}
\end{figure}

A graphical display of the estimated coefficient functions of all methods is presented in Figure~\ref{fig:Fig_7}. The pffr surface for $\beta(t,s)$ is nearly flat, indicating little detected association between wind speed and wind chill once a smooth mean is fit-an expected consequence when spatial endogeneity in the response is unmodeled and the penalty strongly shrinks the bivariate effect. SFoFR reveals a low-amplitude, spatially rough pattern. In contrast, PSFoFR concentrates the signal along a band near $t \approx s$ with the strongest effects in the winter portion of the domain, indicating a predominantly contemporaneous influence of wind speed on wind chill that intensifies in colder months. Off-diagonal structure is weaker, suggesting limited lagged carryover day-to-day once the spatial lag of $\Y$ is instrumented and estimated. This pattern accords with the physical formula for wind chill, in which the (instantaneous) effect of higher wind speed is to lower the perceived temperature at that time. The sharper relief in PSFoFR relative to SFoFR reflects two features of the proposed method: (i) explicit penalization, which stabilizes the bivariate surface and suppresses spurious high-frequency variation; and (ii) IV estimation, which corrects attenuation from the endogenous spatially lagged response, allowing the contemporaneous ridge to emerge more distinctly.

\begin{figure}[!htb]
\centering
\includegraphics[width=9cm]{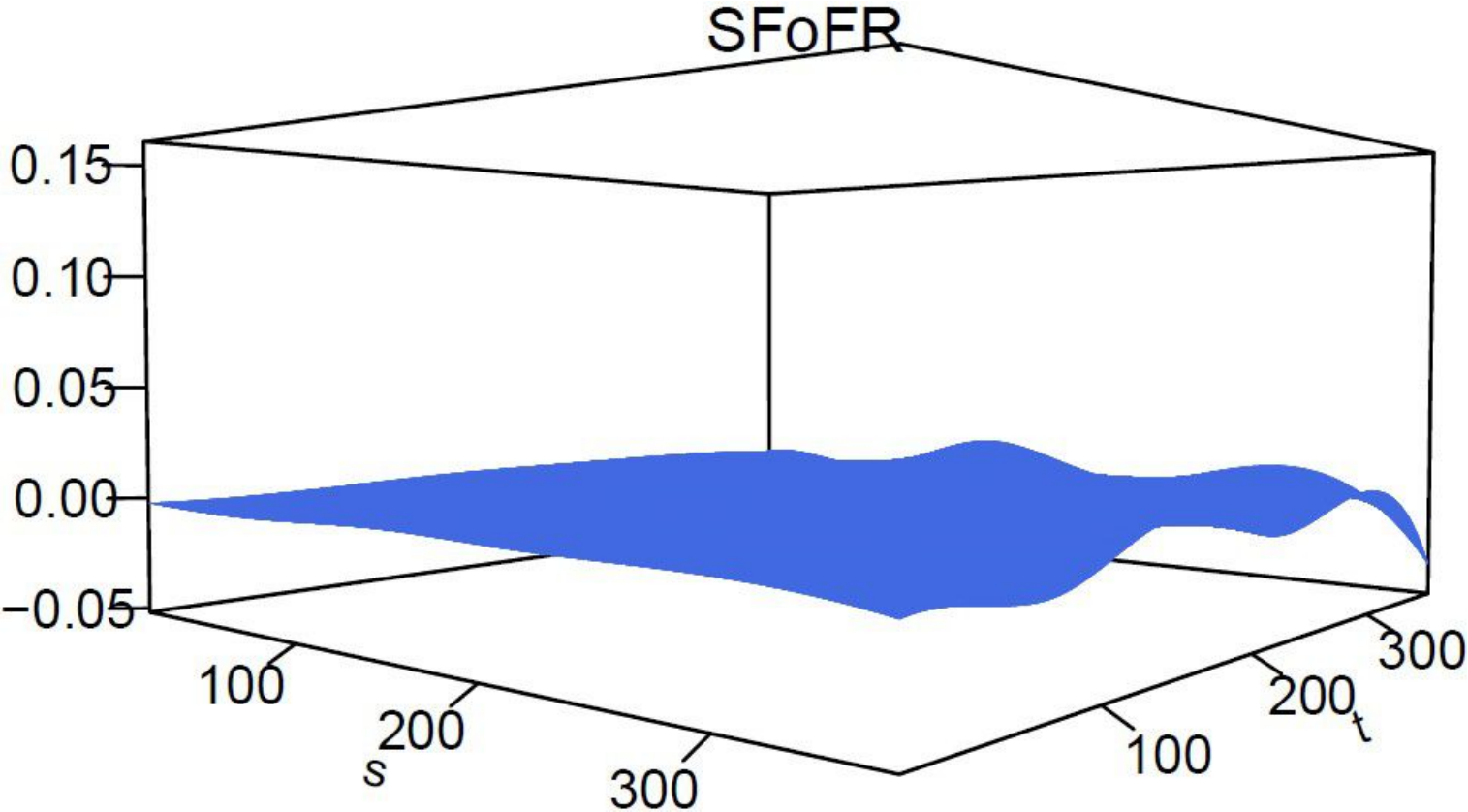}
\quad
\includegraphics[width=9cm]{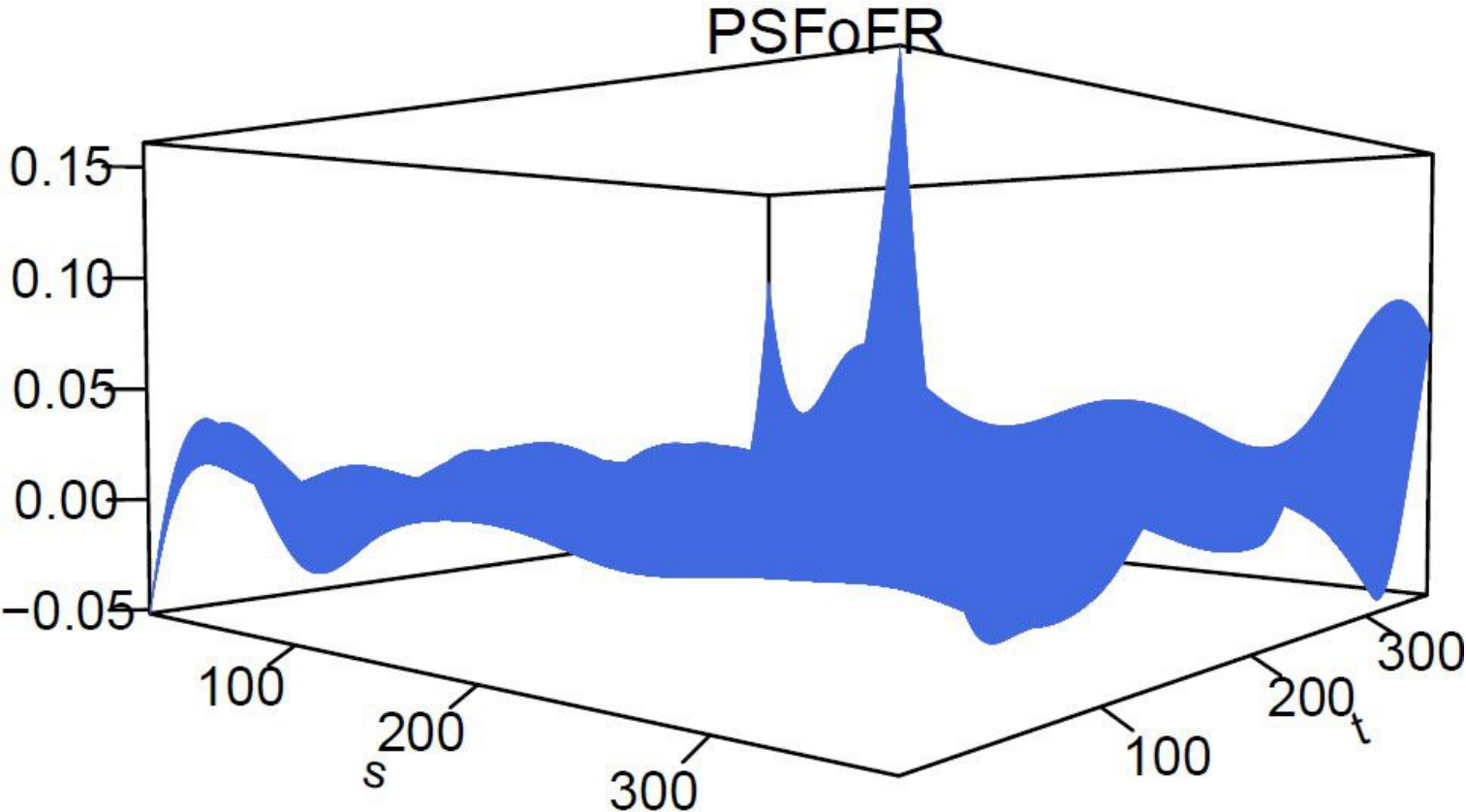}
\caption{Three-dimensional representation of the estimated spatial autocorrelationt function $\rho(t,u)$ by SFoFR (left panel) and PSFoFR (right panel).}\label{fig:Fig_8}
\end{figure}

The estimated autocorrelation functions by the SFoFR and the proposed PSFoFR approaches are presented in Figure~\ref{fig:Fig_8}. The estimated spatial autocorrelation function by SFoFR appears smooth and relatively flat with minimal fluctuations. This suggests that the SFoFR model captures only low-magnitude spatial dependencies and may struggle to detect localized variations in spatial correlation. The lack of prominent peaks or variations indicates potential underfitting, where important spatial dependencies might be missed. However, the estimated spatial autocorrelation function by the proposed PSFoFR method exhibits more pronounced fluctuations and sharper peaks, indicating a higher sensitivity to spatial variations. The presence of strong peaks suggests that the PSFoFR model effectively captures localized spatial dependencies, which were not detected by the SFoFR model.

\section{Conclusion}\label{sec:6}

We propose a novel SFoFR model to address the challenges of modeling spatially correlated functional data. By incorporating a spatial autoregressive structure and penalizing the regression coefficient function using a tensor product of B-splines, our approach effectively captures both functional and spatial dependencies while ensuring smoothness and interpretability. We introduce a PenS2SLS estimator, which balances bias-variance trade-off and mitigates overfitting. The theoretical properties of the proposed estimator, including $\sqrt{n}$-consistency and asymptotic normality, are rigorously established under mild regularity conditions.

The Monte Carlo results confirm that PSFoFR significantly outperforms both SFoFR and pffr by successfully combining penalization and spatial autocorrelation modeling. The penalization in PSFoFR effectively mitigates overfitting, leading to more stable and accurate parameter estimation compared to SFoFR, while the inclusion of spatial dependence provides substantial improvements over pffr, particularly in highly correlated settings. The proposed method achieves the best estimation accuracy, the lowest prediction errors, and the most reliable confidence intervals. The North Dakota weather data analysis further demonstrates the practical advantages of our approach in capturing spatial and functional interactions in empirical settings.

Despite these advancements, our method has some limitations. The selection of smoothing and spatial weighting parameters may be computationally intensive, particularly for large datasets. Additionally, our model assumes a stationary spatial dependence structure, which may not always hold in heterogeneous spatial environments.

Several future research directions include the following. First, development of adaptive smoothing techniques that dynamically adjust penalization parameters in a data-driven manner, improving flexibility and computational efficiency. Second, extending the model to nonstationary spatial processes would allow more realistic modeling of spatially heterogeneous functional data. Third, incorporation of Bayesian hierarchical formulations or deep learning-based spatial functional models could significantly enhance both scalability and predictive performance in high-dimensional applications. Fourth, extending the model to a spatial function-on-functional linear quantile regression framework \cite[see, e.g.,][]{BSS2025} may enable a comprehensive characterization of the conditional distribution of the spatially correlated functional response. Fifth, integrating robust estimation procedures, such as those in \cite{Cai2021} and \cite{BSM2025}, to provide improved resilience against outliers has the potential to ensure more stable and reliable inference in the presence of data contamination.

\section*{Acknowledgment}

We thank the Associate Editor and two reviewers for their thorough reading of our manuscript and for their insightful suggestions and comments, which have significantly contributed to the improvement of our article. This research was financially supported by the Scientific and Technological Research Council of Turkey (TUBITAK) (grant no. 124F096), an Australian Research Council Discovery Project (grant no. DP170102468), and an Australian Research Council Future Fellowship (grant no. FT240100338).

\vspace{-.05in}

\section*{Conflicts of interests/Competing interests}

The authors have no conflicts of interest to declare that are relevant to the content of this article.

\section*{Appendix}

We present the consistency and asymptotic normality of the proposed estimators. We require the following assumptions for the consistency and asymptotic normality of the estimators. 

\begin{itemize}
\item[$A_1$] 
\begin{itemize}
\item[(i)] $w_{ii} = 0$ for all $i$.
\item[(ii)] $\sum_{j=1}^n w_{ij} = 1$ for all $i$.
\item[(iii)] $\Vert \bm{W} \Vert_\infty \leq a_w$ for some $a_w < \infty$.
\item[(iv)] We assume that $\Vert \rho \Vert_\infty < 1 / \bm{W}$. Let $\mathcal{T}$ and $\mathbb{I}_d$ defined as in Remark~\ref{rem1}. Assume that the operator $\mathbb{I}_d - \mathcal{T}$ is invertible and thus $(\mathbb{I}_d - \mathcal{T})^{-1}$ exists. 
\end{itemize}
\item[$A_2$]
\begin{itemize}
    \item[(i)] The true bivariate coefficient functions $ \rho(t,u) \in \mathcal{C}^{p+1}([0,1]^2)$ and $\beta(t,s) \in \mathcal{C}^{p+1}([0,1]^2)$ satisfy:
    \begin{align*}
    \sup_{t,u \in [0,1]} \left| \rho(t,u) - \sum_{\ell=1}^{K_y} \sum_{m=1}^{K_y} \rho_{\ell m} \phi_{\ell}(t) \phi_m(u) \right| &= \mathcal{O} (K_y^{-(p^*+1)}), \\
    \sup_{t,s \in [0,1]} \left| \beta(t,s) - \sum_{\ell=1}^{K_y} \sum_{k=1}^{K_x} b_{\ell k} \phi_{\ell}(t) \psi_k(s) \right| &= \mathcal{O} (K_y^{-(p^*+1)} + K_x^{-(p^*+1)})
    \end{align*}
    where $\{ \phi_{\ell}(t) \}_{\ell=1}^{K_y}$ and $\{ \psi_k(s) \}_{k=1}^{K_x}$ are B-spline bases of degree $p^*$. For cubic B-splines ($p^*=3$), the functions $\rho(\cdot, \cdot)$ and $\beta(\cdot, \cdot)$ must be $\mathcal{C}^4$ and the B-spline basis functions are Lipschitz continuous with a constant.
    \item[(ii)] The matrices $D_t$, $D_u$, and $D_s$ in penalty matrix 
    \begin{equation*}
    \bm{R}(\lambda_{\rho}, \lambda_{\beta}) = 
    \begin{pmatrix}
    \lambda_{\rho} (\bm{\Phi}_u \otimes \bm{D}_t + \bm{D}_u \otimes \bm{\Phi}_t) & 0 \\
    0 & \lambda_{\beta} (\bm{\Psi} \otimes \bm{D}_t + \bm{D}_s \otimes \bm{\Phi}_t)
    \end{pmatrix}.
    \end{equation*}
    are matrices of integrated squared $q^* = 2$\textsuperscript{nd} derivatives of the B-spline basis functions.
\end{itemize}
\item[$A_3$]
\begin{itemize}
    \item[(i)] The number of knots $K_y$ and $K_x$ grows with the sample size $n$ as $K_y = \mathcal{O}(n^{\nu_y})$, $K_x = \mathcal{O}(n^{\nu_x})$, where $\nu_x, \nu_y < \frac{1}{2p^*+3}$.
    \item[(ii)] The penalty parameters $\lambda_{\rho}$ and $\lambda_{\beta}$ decay as $\lambda_{\rho} = o(n^{-1} K_y^{2q^*})$ and $\lambda_{\beta} = o(n^{-1} K_x^{2q^*})$, where $q^{*} \leq p^{*}$.
\end{itemize}
\item[$A_4$]
\begin{itemize}
    \item[(i)] The IV matrix $\bm{Z} = (\bm{Z}_0, \bm{Z}_1, \ldots, \bm{Z}_Q)$ includes columns from $\{ \widetilde{\bm{\psi}}^{(q)} \otimes \bm{\phi}^*, \widetilde{\bm{\phi}} \otimes \bm{\phi}^* \}$ and satisfies:
    \begin{equation*}
    \operatorname{plim}_{n \to \infty} \frac{1}{n} \bm{Z}^\top \bm{\Pi} = \bm{Q}_{Z\Pi}, \qquad \operatorname{rank}(\bm{Q}_{Z\Pi}) = K_y K_x.
    \end{equation*}
    We assume that the minimum eigenvalue of $\bm{Q}_{Z\Pi}$ is positive.
    \item[(ii)] $\operatorname{plim}_{n \to \infty} \frac{1}{n} \bm{Z}^\top \epsilon = 0$, where $\epsilon = \operatorname{vec}\{ \epsilon_i(t_{\iota}) \}$.
\end{itemize}
\item[$A_5$] The error term $\epsilon_i(t)$ is independent of spatial units $i$ conditional on the spatial operator $\mathcal{T}$. All spatial dependence is captured by $\mathcal{T}$ leaving $\epsilon_i(t)$ independently and identically distributed with $\mathbb{E} \{\epsilon_i(t)\} = 0$, $\text{Var} \{\epsilon_i(t) \} = \sigma^2$, and $\mathbb{E} ( \vert \epsilon_i(t) \vert^{4+\delta})$ for some $\delta > 0$.
\end{itemize}

Assumption $A_1$ is essential for the well-posedness of spatial dependencies, ensuring the identifiability of the spatial autoregressive term and preventing singularities or instability in the estimation procedure. This condition guarantees that spatial interactions among functional observations do not lead to ill-posed problems, thereby allowing for a consistent estimation \citep[see, e.g.,][]{Kelejian1998, Lee2004, BSGARC2024}. Assumptions $A_2$ and $A_3$ play a fundamental role in defining the regression coefficient function and the spatial autocorrelation function within an infinite-dimensional functional space. These conditions justify the application of basis function expansions and penalization techniques, which serve to control the roughness of the estimated coefficients. This regularization prevents overfitting while ensuring stable numerical estimation. In particular, Assumption $A_2~(i)$ extends the results of \cite{Claeskens2009}, who demonstrated that for any univariate function $f(t) \in \mathcal{C}^p$, the approximation error satisfies $\vert f(t) - \sum_{k=1}^K f_k \upsilon_k(t) \vert = o(K^{p^*+1})$ where $\{\upsilon_k(t)\}$ denotes a spline basis and $\{f_1,\dots, f_K\}$ are the associated basis coefficients. Although this result holds for univariate functions, the same error bound extends to tensor-product spline approximations, thus validating Assumption $A_2~(i)$  in the functional setting \citep[see, e.g.,][]{Mobler2009, Belloni2015, Lyche2018}. Assumption $A_4$ is crucial to establish the consistency and asymptotic normality of the finite-dimensional PenS2SLS estimator $\widehat{\bm{\theta}}$ \citep[see, e.g.,][]{Kelejian1998}. Finally, Assumption $A_5$ is fundamental for proving the $\sqrt{n}$-consistency and asymptotic normality of the proposed PenS2SLS estimator. This condition ensures that the instrumental variable approach effectively mitigates endogeneity issues arising from the spatially lagged functional response, thereby enabling valid inference under an asymptotic Gaussian process framework.

\begin{proof}[Proof of Theorem~\ref{th:1}]
The estimators $\widehat{\rho}(t,u)$ and $\widehat{\beta}(t,s)$ are linear transformations of the finite-dimensional basis coefficients $\widehat{\bm{\theta}} = (\widehat{\widetilde{\bm{\rho}}}^\top, \widehat{\widetilde{\bm{b}}}^\top)^\top$. Let $\bm{\theta} = (\widetilde{\bm{\rho}}^\top, \widetilde{\bm{b}}^\top)^\top$ denote the true coefficients in the tensor product of the B-spline bases. We define the total estimation error as:
\begin{align*}
\widehat{\rho}(t,u) - \rho(t,u) &= \{\widehat{\rho}(t,u) - \rho_K(t,u) \} + \{\rho_K(t,u) - \rho(t,u) \}, \\
\widehat{\beta}(t,s) - \beta(t,s) &= \{\widehat{\beta}(t,s) - \beta_K(t,s) \} + \{\beta_K(t,s) - \beta(t,s) \},
\end{align*}
where $\rho_K(t,u) = \sum_{\ell=1}^{K_y} \sum_{m=1}^{K_y} \rho_{\ell m} \phi_{\ell}(t) \phi_m(u)$ and $\beta_K(t,s) = \sum_{\ell=1}^{K_y} \sum_{k=1}^{K_x} b_{\ell k} \phi_{\ell}(t) \psi_k(s)$ are the B-spline projections of $\rho(t,u)$ and $\beta(t,s)$, respectively. Here, the errors $\{\rho_K(t,u) - \rho(t,u) \}$ and $\{\beta_K(t,s) - \beta(t,s) \}$ are the approximation errors, while $\{\widehat{\rho}(t,u) - \rho_K(t,u) \}$ and $\{\widehat{\beta}(t,s) - \beta_K(t,s) \}$ are the estimation errors.

Under Assumption $A_2$, the tensor product B-spline approximation satisfies:
\begin{align*}
\sup_{t,u} \left| \rho(t,u) - \rho_K(t,u) \right| &= \mathcal{O} (K_y^{-(p^*+1)}), \\
\sup_{t,s} \left| \beta(t,s) - \beta_K(t,s) \right| &= \mathcal{O} (K_y^{-(p^*+1)} + K_x^{-(p^*+1)}).
\end{align*}
By $A_3~(i)$, $K_y = \mathcal{O}(n^{\nu_y})$, $K_x = \mathcal{O}(n^{\nu_x})$, with $\nu_y, \nu_x < \frac{1}{2p^*+3}$. Thus:
\begin{align*}
K_y^{-(p^*+1)} &= \mathcal{O} (n^{1/2 - \nu_y (p^*+1)}) \to 0, \\
K_x^{-(p^*+1)} &= \mathcal{O} (n^{1/2 - \nu_x (p^*+1)}) \to 0.
\end{align*}
Hence, the approximation error terms vanish asymptotically.

From Model~\eqref{eq:sfofrm} and Remark~\ref{rem1}, under $A_1$, the reduced form is:
\begin{equation*}
\Y_i(t) = (\mathbb{I}_d - \mathcal{T})^{-1} \left\{ \int_0^1 \X_i(s) \beta(t,s) \, ds + \epsilon_i(t) \right\},
\end{equation*}
Substituting the B-spline approximations $\rho_K(t,u)$ and $\beta_K(t,s)$, we rewrite the model in terms of basis coefficients:
\begin{equation*}
\text{vec}(\widetilde{\Y}) = \bm{\Pi} \bm{\theta} \epsilon + \text{vec}(\text{A}_{\text{err}}),
\end{equation*}
where $\text{vec}(\text{A}_{\text{err}})$ represents the approximation error $\rho(t,u) - \rho_K(t,u)$ and $\beta(t,s) - \beta_K(t,s)$. By $A_2$, $\Vert \text{vec}(\text{A}_{\text{err}}) \Vert = o_p(1)$. Thus, the PenS2SLS estimator is:
\begin{equation*}
\widehat{\bm{\theta}} = \left\lbrace\widetilde{\bm{\Pi}}^\top + \bm{R}(\lambda_\rho, \lambda_\beta) \right\rbrace^{-1} \widehat{\bm{\Pi}}^\top \text{vec}(\widetilde{\Y}).
\end{equation*}

By substituting $\text{vec}(\widetilde{\Y}) = \bm{\Pi} \bm{\theta} + \epsilon$ into $\widehat{\bm{\theta}}$ we have:
\begin{equation}\label{eq:A1}
\widehat{\bm{\theta}} = \left\lbrace \bm{\Pi}^\top \bm{\Pi} + \bm{R}(\lambda_\rho, \lambda_\beta) \right\rbrace^{-1} (\bm{\Pi}^\top \bm{\Pi} \bm{\theta} + \bm{\Pi}^\top \epsilon)
\end{equation}
Subtracting $\bm{\theta}$ from both sides of~\eqref{eq:A1} yields:
\begin{align}
\widehat{\bm{\theta}} - \bm{\theta} &= \left\lbrace \bm{\Pi}^\top \bm{\Pi} + \bm{R}(\lambda_\rho, \lambda_\beta) \right\rbrace^{-1} (\bm{\Pi}^\top \bm{\Pi} \bm{\theta} + \bm{\Pi}^\top \epsilon) - \bm{\theta}, \nonumber \\
&= \left\lbrace \bm{\Pi}^\top \bm{\Pi} + \bm{R}(\lambda_\rho, \lambda_\beta) \right\rbrace^{-1} \bm{\Pi}^\top \epsilon + \left\lbrace \bm{\Pi}^\top \bm{\Pi} + \bm{R}(\lambda_\rho, \lambda_\beta) \right\rbrace^{-1} \bm{\Pi}^\top \bm{\Pi} \bm{\theta} - \bm{\theta}. \label{eq:A2}
\end{align}
By factoring out $\left\lbrace \bm{\Pi}^\top \bm{\Pi} + \bm{R}(\lambda_\rho, \lambda_\beta) \right\rbrace^{-1}$, we have:
\begin{equation*}
\widehat{\bm{\theta}} - \bm{\theta} = \left\lbrace \bm{\Pi}^\top \bm{\Pi} + \bm{R}(\lambda_\rho, \lambda_\beta) \right\rbrace^{-1} \bm{\Pi}^\top \epsilon + \left\lbrace \bm{\Pi}^\top \bm{\Pi} + \bm{R}(\lambda_\rho, \lambda_\beta) \right\rbrace^{-1} \left[ \bm{\Pi}^\top \bm{\Pi} \bm{\theta} - \left\lbrace \bm{\Pi}^\top \bm{\Pi} + \bm{R}(\lambda_\rho, \lambda_\beta) \right\rbrace \bm{\theta} \right]
\end{equation*}
Since $\bm{\Pi}^\top \bm{\Pi} \bm{\theta} - \left\lbrace \bm{\Pi}^\top \bm{\Pi} + \bm{R}(\lambda_\rho, \lambda_\beta) \right\rbrace \bm{\theta} = - \bm{R}(\lambda_\rho, \lambda_\beta) \bm{\theta}$, we obtain:
\begin{equation*}
\widehat{\bm{\theta}} - \bm{\theta} = \left\lbrace \bm{\Pi}^\top \bm{\Pi} + \bm{R}(\lambda_\rho, \lambda_\beta) \right\rbrace^{-1} \bm{\Pi}^\top \epsilon - \left\lbrace \bm{\Pi}^\top \bm{\Pi} + \bm{R}(\lambda_\rho, \lambda_\beta) \right\rbrace^{-1} \bm{R}(\lambda_\rho, \lambda_\beta) \bm{\theta}
\end{equation*}

Under $A_3~(ii)$, $\lambda_{\rho} = o(n^{-1} K_y^{2q^*})$ and $\lambda_{\beta} = o(n^{-1} K_x^{2q^*})$. Thus, 
\begin{equation*}
\frac{1}{n} \bm{R}(\lambda_\rho, \lambda_\beta) = \frac{1}{n}
\begin{pmatrix}
\lambda_{\rho} (\bm{\Phi}_u \otimes \bm{D}_t + \bm{D}_u \otimes \bm{\Phi}_t) & 0 \\
0 & \lambda_{\beta} (\bm{\Psi} \otimes \bm{D}_t + \bm{D}_s \otimes \bm{\Phi}_t)
\end{pmatrix} \xrightarrow{p} 0.
\end{equation*}
This implies
\begin{equation*}
\left\lbrace \frac{1}{n} \bm{\Pi}^\top \bm{\Pi} \frac{1}{n} \bm{R}(\lambda_\rho, \lambda_\beta) \right\rbrace^{-1} = \mathcal{\bm{V}}^{-1}, \quad \text{where} ~~ \mathcal{\bm{V}} = \operatorname{plim}_{n \to \infty} \frac{1}{n} \bm{\Pi}^\top \bm{\Pi}.
\end{equation*}

By $A_4$ and $A_5$, the design matrix and IVs satisfy $\frac{1}{n} \bm{\Pi}^\top \bm{\Pi} \xrightarrow{p} \mathcal{\bm{V}}$ and $\frac{1}{n} \bm{\Pi}^\top \epsilon \xrightarrow{p}0$. Thus, using Slutsky's theorem, we obtain:
\begin{equation*}
\widehat{\bm{\theta}} - \bm{\theta} = \mathcal{\bm{V}}^{-1} 0 + o_p(1) = o_p(1).
\end{equation*}
In other words, the PenS2SLS estimator of $\bm{\theta}$ is a consistent estimator, that is, $\widehat{\bm{\theta}} \xrightarrow{p} \bm{\theta}$.

Now, expanding the PenS2SLS estimator $\widehat{\bm{\theta}}$, we obtain:
\begin{equation*}
\sqrt{n} (\widehat{\bm{\theta}} - \bm{\theta}) = \left\lbrace \frac{1}{n} \bm{\Pi}^\top \bm{\Pi} + \frac{1}{n} \bm{R}(\lambda_\rho, \lambda_\beta) \right\rbrace^{-1} \frac{1}{\sqrt{n}} \bm{\Pi}^\top \epsilon - \left\lbrace \frac{1}{n} \bm{\Pi}^\top \bm{\Pi} + \frac{1}{n} \bm{R}(\lambda_\rho, \lambda_\beta) \right\rbrace^{-1} \frac{1}{\sqrt{n}} \bm{R}(\lambda_\rho, \lambda_\beta) \bm{\theta}.
\end{equation*}
Under $A_3~(ii)$, $\frac{1}{\sqrt{n}} \bm{R}(\lambda_\rho, \lambda_\beta) \bm{\theta} \xrightarrow{p}0$, so
\begin{equation*}
\sqrt{n} (\widehat{\bm{\theta}} - \bm{\theta}) = \left( \frac{1}{n} \bm{\Pi}^\top \bm{\Pi} \right)^{-1} \frac{1}{\sqrt{n}} \bm{\Pi}^\top \epsilon + o_p(1).
\end{equation*}
By $A_4~(ii)$, $A_5$, and Theorem 3 of \cite{Kelejian1998}, the score satisfies:
\begin{equation*}
\frac{1}{\sqrt{n}} \bm{\Pi}^\top \epsilon \xrightarrow{p} \mathcal{N}(0, \sigma^2 \mathcal{\bm{V}}^{-1}).
\end{equation*}
Thus, the PenS2SLS estimator $\widehat{\bm{\theta}}$ has an asymptotic normal distribution:
\begin{equation}\label{eq:A3}
\sqrt{n} (\widehat{\bm{\theta}} - \bm{\theta}) \xrightarrow{d} \mathcal{N} (0, \sigma^2 \mathcal{\bm{V}}^{-1}).
\end{equation}

The functional estimators $\widehat{\rho}(u,t)$ and $\widehat{\beta}(t,s)$ are linear transformations of the finite-dimensional parameter estimates $\widetilde{\widehat{\bm{\rho}}}$ and $\widetilde{\widehat{\bm{b}}}$, respectively:
\begin{align*}
\widehat{\rho}(u,t) &= \{\bm{\phi}^\top(t) \otimes \bm{\phi}^\top (u) \} \widetilde{\widehat{\bm{\rho}}} \\ 
\widehat{\beta}(t,s) &= \{\bm{\phi}^\top(t) \otimes \bm{\psi}^\top (s) \} \widetilde{\widehat{\bm{b}}}.
\end{align*}
By the continuous mapping theorem, which states that if a sequence of random vectors $\bm{X}_n$ converges in distribution $\bm{X}$ and $g$ is a continuous function, then $g(\bm{X}_n)$ converges in distribution to $g(\bm{X})$, we have~\eqref{eq:A3}. The functional estimators $\widehat{\rho}(u,t)$ and $\widehat{\beta}(t,s)$ are continuous functions of $\widehat{\bm{\theta}}$. Spectically, for fixed $(t,u)$ and $(t,s)$, we defined the following linear operators:
\begin{align*}
L_\rho (t,u) &= \bm{\phi}^\top(t) \otimes \bm{\phi}^\top (u) \\ 
L_\beta (t,s) &= \bm{\phi}^\top(t) \otimes \bm{\psi}^\top (s).
\end{align*}
Then, we have
\begin{align*}
\sqrt{n} \{\widehat{\rho}(t,u) - \rho(t,u) \} &= L_\rho (t,u) \sqrt{n} (\widetilde{\widehat{\bm{\rho}}} - \widetilde{\bm{\rho}}), \\
\sqrt{n} \{\widehat{\beta}(t,s) - \beta(t,s) \} &= L_\beta (t,s) \sqrt{n} (\widetilde{\widehat{\bm{b}}} - \widetilde{\bm{b}})
\end{align*}
By the continuous mapping theorem, since $\sqrt{n} (\widetilde{\widehat{\bm{\rho}}} - \widetilde{\bm{\rho}})$ and $\sqrt{n} (\widetilde{\widehat{\bm{b}}} - \widetilde{\bm{b}})$ converge in distribution to normal random vectors, $\sqrt{n} \{\widehat{\rho}(t,u) - \rho(t,u) \}$ and $\sqrt{n} \{\widehat{\beta}(t,s) - \beta(t,s)\}$ also converge in distribution random variables.

Since the asymptotic covariance of $\sqrt{n} (\widetilde{\widehat{\bm{\rho}}} - \widetilde{\bm{\rho}})$ and $\sqrt{n} (\widetilde{\widehat{\bm{b}}} - \widetilde{\bm{b}})$ is $\sigma^2 \mathcal{\bm{V}}^{-1}$, we obtain the asymptotic covariance of $\sqrt{n} \{\widehat{\rho}(t,u) - \rho(t,u) \}$ and $\sqrt{n} \{\widehat{\beta}(t,s) - \beta(t,s) \}$ as follows:
\begin{align}
\bm{\Sigma}_\rho (t,u; t^\prime, u^\prime) &= \sigma^2 L_\rho (t,u) \mathcal{\bm{V}}^{-1} L_\rho^\top (t^\prime, u^\prime), \label{eq:A4} \\
\bm{\Sigma}_\beta (t,s; t^\prime, s^\prime) &= \sigma^2 L_\beta (t,s) \mathcal{\bm{V}}^{-1} L_\beta^\top (t^\prime, s^\prime). \label{eq:A5}
\end{align}
Substituting $L_\rho (t,u) = \bm{\phi}^\top(t) \otimes \bm{\phi}^\top (u)$ and $L_\beta = \bm{\phi}^\top(t) \otimes \bm{\psi}^\top (s)$ in~\eqref{eq:A4} and~\eqref{eq:A5}, respectively, we obtain
\begin{align*}
\bm{\Sigma}_\rho (t,u; t^\prime, u^\prime) &= \sigma^2 \{\bm{\phi}^\top(t) \otimes \bm{\phi}^\top (u) \} \mathcal{\bm{V}}^{-1} \{ \bm{\phi}(t^\prime) \otimes \bm{\phi} (u^\prime) \}, \\
\bm{\Sigma}_\beta (t,s; t^\prime, s^\prime) &= \sigma^2 \{ \bm{\phi}^\top \otimes \bm{\psi}^\top (s) \} \mathcal{\bm{V}}^{-1} \{ \bm{\phi}^\top(t^\prime) \otimes \bm{\psi}^\top (s^\prime) \}.
\end{align*}

Under assumptions $A_1$--$A_5$, the linear mappings from the finite-dimensional coefficients to the functional space preserve asymptotic normality. Specifically:
\begin{align*}
\sqrt{n} \{\widehat{\rho}(t,u) - \rho(t,u) \} \xrightarrow{d} \mathcal{GP} \{0, \bm{\Sigma}_\rho (t,u; t^\prime, u^\prime) \}, \\
\sqrt{n} \{\widehat{\beta}(t,s) - \beta(t,u) \} \xrightarrow{d} \mathcal{GP} \{0, \bm{\Sigma}_\beta (t,s; t^\prime, s^\prime) \}.
\end{align*}
where $\mathcal{GP}\{\cdot,\cdot\}$ denotes the Gaussian process.
\end{proof}

\bibliographystyle{apalike}
\bibliography{sfof.bib}

\end{document}